\documentclass[printer]{aa}  

\usepackage{subfigure}
\usepackage{multirow}
\usepackage{tablefootnote}

\usepackage{graphicx}
\usepackage{txfonts}
\begin{document}

\title{On the stability around Chariklo and the confinement of its rings}
\titlerunning{Chariklo and the confinement of its rings}
\authorrunning{Giuliatti~Winter et al.}

\author{S.M. {Giuliatti~Winter},
          \inst{1}\thanks{\email{giuliatti.winter@unesp.br}}
          G. {Madeira},\inst{2,1}
T. {Ribeiro},\inst{1}
O.C. {Winter},\inst{1}
G.O. {Barbosa},\inst{1,3}
\and G. {Borderes-Motta}\inst{4}
          }

   \institute{Grupo de Din\^amica Orbital \& Planetologia, S\~ao Paulo State University - UNESP, Av. Ariberto Pereira da Cunha, 333, Guaratinguet\'a SP, 12516-410, Brazil\\
   \and
   {Universit\'e de Paris, Institut de Physique du Globe de Paris, CNRS F-75005 Paris, France}\\
   \and
Eldorado Research Institute, Av. Prof. Alan Turing, 275 - Cidade Universit\'aria, Campinas - SP, Brazil\\ \and 
    Swedish Institute of Space Physics -IRF, Box 812, SE-98128 Kiruna, Sweden
    }

   \date{Received XXXX; accepted XXXX}

% \abstract{}{}{}{}{} 
% 5 {} token are mandatory
 
  \abstract
{Chariklo has two narrow and dense rings, C1R and C2R, located at 391~km and 405~km, respectively.  } 
   {In the light of new stellar occultation data, we study the stability around Chariklo. We also analyse three confinement mechanisms, to prevent the spreading of the rings, based on shepherd satellites in resonance with the edges of the rings.  }
   {This study is made through a set of numerical simulations and the  Poincar\'e surface of section technique.}
   {From the numerical simulation results we verify that, from the  current parameters referring to the shape of Chariklo,  the inner edge of the stable region is much closer to Chariklo than the rings. 
The Poincar\'e surface of sections allow us to identify the first kind periodic and quasi-periodic orbits, and also the resonant islands corresponding to the 1:2, 2:5, and 1:3 resonances. We construct a   map of  $a_{eq}$ versus $e_{eq}$ space which gives the location and width of the stable region and the 1:2, 2:5, and 1:3 resonances.  }
   {We found that the first kind periodic orbits family can be responsible for a stable region whose location and size meet that of C1R, for specific values of the ring particles' eccentricities. However, C2R is located in an unstable region if the width of the ring is assumed to be about 120~m. After analysing  different systems we propose that the best confinement mechanism is composed of three satellites, two of them shepherding the inner edge of C1R and the outer edge of C2R, while the third satellite would be  trapped in the 1:3  resonance. }


   \maketitle
\defcitealias{BragaRibas2014}{BR14}
\defcitealias{Morgado2021}{M21}

%
%-------------------------------------------------------------------

\section{Introduction}

The amazing  rings of Saturn were discovered in the seventeen century by  Galileo Galilei, although he died before  knowing he had discovered a unique  planetary ring system. Only three centuries later,  the   rings around  the giant planets Jupiter, Uranus and Neptune  were revealed by the  spacecraft Voyager~I and II and by stellar occultations.

A new class of objects sheltering a ring system was discovered in  2014 by Braga~Ribas and colleagues [2014, BR14 hereafter]{BragaRibas2014} through  stellar occultation. They  discovered a ring system around the largest   Centaur (10199)~Chariklo. This Centaur object could  originate in the trans-Neptunian region and could be deflected to the Centaur region probably due to a close encounter with Neptune within the last 20~Myr \citep{Wood2017}.

The two  dense rings, 2013C1R and 2013C2R, in orbit around Charilko   are  very narrow rings  with widths of about 7~km and 3~km and optical depths of 0.4 and 0.06, respectively \citepalias{BragaRibas2014}. They are located very close to Chariklo, their orbital radii are 391~km and 405~km.  \citetalias{BragaRibas2014} speculated three scenarios  for the  origin of the two rings, one of them relies on  a collision between  a satellite onto the surface of Chariklo. This collision could release material from Chariklo and forms the rings, or the impactor satellite could  be destroyed originating the rings.  

Regarding the composition of the rings, \cite{Duffard2014} obtained  that silicates, tholins and water ice  may be present in C1R and C2R, while   \cite{Sicardy2020} claimed that the presence of icy water,   shown in the spectrum of the rings, may be caused by Chariklo.

A set of five stellar occultations presented in  \cite{Leiva2017} between 2013 and 2016 helped to constrain the size and shape of  Chariklo. The shape of an object can help to analyse the dynamical behaviour that nearby particles  can perform, and in this particular case,   could help  to understand  the  origin and evolution of the rings. They considered four possible models for Chariklo: a sphere with radius  equal to $129$~km, a MacLaurin spheroid with about  $a = b = 143$~km and $c=96$~km, a triaxial ellipsoid with $a= 157$~km, $b = 132$~km and $c= 102$~km, and  a Jacobi ellipsoid with $a= 157$~km, $b= 139$~km and $c= 86$~km, where $a$, $b$ and $c$ are the  semi-axes.  With the  derived mass range  for  Chariklo of  $6-8 \times 10^{18}$~kg, they pointed out  that the  1:3 resonance between the  rotation of Chariklo and the orbital motion of the particles  is at $408 \pm 20$~km, close to the location of the rings. 

 A  paper  by \cite{Morgado2021} -- \citetalias{Morgado2021} hereafter -- presents new stellar occultation data obtained  between 2017 and 2020. These new data helped to improve Chariklo and the rings parameters. The parameters of C1R and C2R can be consulted in Table \ref{tab:char}. They also concluded that these rings may contain particles larger than $1~\mu$m in size. An important result concerns  the shape of Chariklo, from these stellar occultations data  Chariklo is consistent with a triaxial ellipsoid with semi-axes  $a = 143.8$~km, $b=135.2$~km and $c=99.1$~km  \citepalias{Morgado2021}.

The relation between the inner ring of Chariklo and the 1:3 spin-orbit resonance (resonance between Chariklo's rotation period and the mean motion of the particles) was explored in \cite{madeira2021} by assuming Chariklo as a spherical body with a mass anomaly at its equator. Such an assumption is based on observational data that suggests the presence of topographic features in the Centaur \citep{Sicardy2019}. Through a set of Poincar\'e surface of sections, \cite{madeira2021} obtain that the non-spherical shape of Chariklo is responsible for an unstable region extending from its surface to an orbital radius of $\sim$320~km, far inside the ring system. Despite the proximity between the 1:3 spin-orbit resonance  and the inner ring, the authors verify that such resonance would be responsible for eccentricities $\sim 10^{-2}$,  higher than expected for the ring particles. Their results show that Chariklo rings are  probably associated with first kind orbits.

Several papers analysed the rings dynamics.  \cite{ElMoutamid2014} and \cite{Melita2017} discussed the possibility that the ring was formed from the disruption of an old satellite, which can occur if the satellite crosses the Roche limit of Chariklo. The location of the Roche limit depends on the physical parameters of both objects, and for this to match the current locations of the rings, the satellite must be porous, with a density much lower than Chariklo. In addition, the satellite must have a minimum radius of $\sim7$~km \citep{Melita2017}. However, a problem with this hypothesis is the absence of mechanisms to bring the satellite to this limit. Tidal dissipation, the primary mechanism that could be responsible for this, is not a plausible option. Chariklo's corotation radius is within the locations of the rings. Therefore a satellite beyond the Roche limit would migrate farther away from Chariklo unless it rotated much slower in the past.

Another possibility ruled out by \cite{Melita2017} is that the disruption is caused by a destructive impact of an old satellite with an external projectile, because the estimated timescale for such an event to occur is longer than Chariklo's lifetime. The authors also analysed the impactor flux in the Chariklo region and conclude that the ring formation due to an impact of a projectile with Chariklo is improbable.

\cite{Pan2016} proposed that the close encounter that brought Chariklo to the Centaur region would enormously increase its temperature, being responsible for the sublimation of CO material. They estimated that Chariklo re-accretes part of this material, while the rest settles in the equation plane after multiple collisions. The latter has dust sizes and mechanically sticks together, forming a material disk that spreads out and forms the Chariklo system. More data on the Chariklo composition are needed to verify if such a process can produce the amount of material observed in the rings.

\cite{Hyodo2016} propose that the  ring material  may have formed by the partial tidal disruption of Chariklo during an extreme close encounter. Their model assumed a differenciated Chariklo with an ice mantle and it was analysed using hydrodynamic simulations. According to them, the planet's tidal effects instantly remove material from Chariklo's surface. Over time, such material settles in the equatorial plane mainly within the Roche limit and viscously spreads out. Finally, the material outside the Roche limit accretes into moons that destructively collide with each other, forming the rings and shepherd satellites \citep{Hyodo2015}. This model has the  advantage of explaining the ice composition of the rings \citep{Duffard2014}, while   it does not require the presence of an old satellite. However, it does require extremely rare close encounters \citep{Araujo2016,Wood2017}. Furthermore, Chariklo composition is uncertain and more data is needed to verify  the most  plausible mechanism for the formation of Chariklo system.

  While several relevant  works analysed  possible scenarios for the  origin of  Chariklo rings, our focus in this work is to analyse the stability of the region face to the new parameters obtained for Chariklo \citepalias{Morgado2021} in order to verify if the rings are located in a stable region and/or close to a resonance. This analysis was performed through a set of numerical simulations and also  through the powerful technique of the  Poincar\'e  surface of section. Furthermore, we explore several confinement models for constrain these rings.

Our paper is divided into 6  sections. Section~\ref{dynamical} deals with the dynamical system and also presents the results obtained from the numerical simulations, while in section~\ref{periodicorbit}  the periodic and quasi-periodic orbits are studied in light of the Poincar\'e surface of sections. In section~\ref{locationrings} we analyse the location of the rings in the $a_{eq}$ versus $e_{eq}$ map.  In section~\ref{secconfi} we discusses  different confinement models to constrain the rings.  Conclusions on this work is presented in section~\ref{discussion}.
\begin{table}
\centering
\caption{The orbital radii, the width and the optical depth of both rings, C1R and C2R, are shown here. This table  also includes the semi-axes ($a, b, $ and $c$), the mass and  the rotation period  of Chariklo (\citetalias{BragaRibas2014}; \cite{Leiva2017}; \citetalias{Morgado2021}).}
\label{tab:char}
{\label{tbinit}}
\begin{tabular}{lcc} \hline \hline
                             & C1R                         & C2R                      \\
Orbital radii (km)           & 385.9                       & 399.8                    \\
Width (km)                   & 4.8-9.1                     & 0.117                    \\
Optical depth                & 0.4                         & 0.06                     \\ \hline
                             & \multicolumn{2}{c}{Chariklo}                           \\
${\rm a \times b \times c}$ (km) & \multicolumn{2}{c}{143.8 $\times$ 135.2 $\times$ 99.1} \\
Mass (kg)                    & \multicolumn{2}{c}{${\rm 6.3 \times 10^{18}}$}         \\
Rotation period (h)          & \multicolumn{2}{c}{7.004}       \\ \hline  \hline                    
\end{tabular}
\end{table}

\section{Dynamical System} \label{dynamical}
The equations of motion of a massless particle around Chariklo, considering a body-fixed frame (O$xy$) uniformly rotating with the same spin period of Chariklo,  can be given by \citep{HU2004} 
\begin{equation}
\ddot{x} - 2 \omega \dot{y} = \omega ^{2} x + U_{x} ,
\label{eq:movx}
\end{equation}
and
\begin{equation}
\label{eq:movy}
\ddot{y} + 2 \omega \dot{x} = \omega ^{2} y + U_{y}
\end{equation}
where  $U_{x}$ and $U_{y}$ are the gravitational potential partial derivatives for Chariklo and $\omega$ is its spin velocity. Here, we assume Chariklo represented by a second degree and order gravity field. The oblateness ($C_{20}$) and  the ellipticity ($C_{22}$) are gravitational potential coefficients whose values can be computed from the ellipsoidal semi-axes $(a,b,c)$ as \citep{balmino1994}
\begin{equation}
C_{20}= (2c^2-a^2-b^2)/10{R_e}^2
\label{eq:c20}
\end{equation}
and 
\begin{equation}
C_{22}=(a^2-b^2)/20{R_e}^2 
\label{eq:c22}
\end{equation}
where $a>b>c$ and $R_e=(a\,b\,c)^{1/3}$. The gravitational potential can be expressed as \citep{HU2004}
\begin{equation}
U(x,y)=\frac{\mu}{r}\left( 1 - \left(\frac{R_{e}}{r} \right)^{2} \left[ \frac{C_{20}}{2}-\frac{3 C_{22}}{r^{2}}(x^{2}-y^{2})\right] \right),
\label{eq:U}
\end{equation}
where $\mu$ is the gravity parameter and $r=\sqrt{x^2+y^2}$. Note that the physical major axis of the body is aligned with the O$x$ axis of the rotating system.

In this section the system (Chariklo-ring particle) described by Equations (1) and (2) will be analysed into two different ways: i) through a set of  numerical  simulation and ii) through the  technique of Poincar\'e surface of sections.

\subsection{Numerical Results}
\begin{figure*}
\center
\includegraphics[width=2.0\columnwidth]{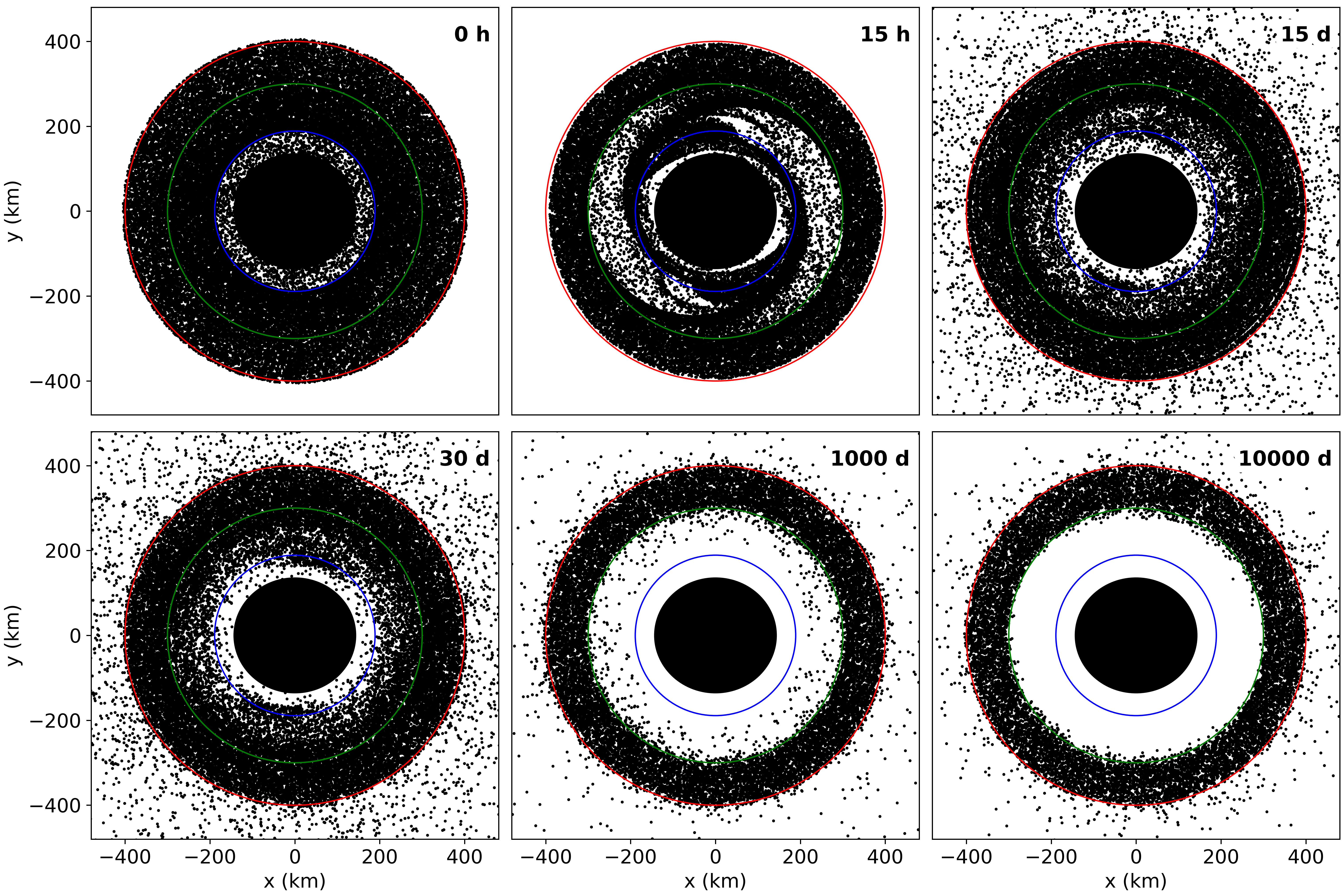}
\caption{These snapshots show the orbital evolution (in the rotational frame with Chariklo) of a set of test  particles under the effects of $C_{20}$ and $C_{22}$, the  gravitational coefficients of Chariklo (black circle). The blue circle shows the location of the corotation radius, while  the green and red circles show the semi-major axes of the 1:2 and 1:3 resonances, respectively.  }
\label{anelpart}
\end{figure*}

 Our first goal is to find the inner edge of the stable region around Chariklo. For that  we performed   numerical simulations using a disc of non-interacting  massless particles moving  around Chariklo. In the gravitational potential of Chariklo we take into account the contributions of its oblateness ($J_2$) and  ellipticity ($C_{22}$). After some changes to include the $C_{22}$ contribution in the equations of motion \citep{celletti2017dynamical}, a N-body simulation in the Rebound package \citep{rein2012rebound} with the IAS15 integrator \citep{rein2015ias15} was used to numerically integrate the system.

The particles were distributed with random values of the true longitude between $0^{\circ} - 360^{\circ}$ from the equivalent radius ($R_{eq})$ of the central body ($\approx$ 124~km) to the position of its main ring ($\approx$ 400~km). Knowing that the inner edge of the  disc suffers large  perturbations due to the  azimuthal asymmetry of the central body, the  amount of initial particles  in this region is low, since  most of them will be ejected or will collide with the central body, consuming unnecessary computing resources. This region that extends from the $R_{eq}$ to the  corotational radius \citep[$\approx$ 189~km,][]{Sicardy2019} has a set of  1000~particles, while  the remaining part  of the disc is composed by 40000~particles. 

All adopted parameters of Chariklo are given in Table~1. A  collision is detected when the orbital radius of the particle  is smaller than the equivalent radius of Chariklo, while those particles with  a semi-major axis  larger  than five times the orbital radius of the main ring are removed from the system. The numerical simulation  has been carried out for 10000~years.

%The particles were distributed with random values of the true longitude between $0^{\circ} - 360^{\circ}$ from the equivalent radius ($R_{eq})$ of the central body ($\approx$ 124~km) to the position of its main ring ($\approx$ 400~km). Knowing that the inner edge of the  disk suffers large  perturbations due to the  azimuthal asymmetry of the central body, the  amount of initial particles  in this region is low, because   most of them will be ejected or collide with the central body, consuming unnecessary computing resources. This region that extends from the $R_{eq}$ to the  corotational radius \citep[$\approx$ 189~km,][]{Sicardy2019} has a set of  1,000 particles, while  the remaining part  of the disk is composed of 40,000 particles. 

%All adopted parameters for Chariklo are given in Table~1. A  collision is detected when the orbital radius of the particle  is smaller than the equivalent radius of Chariklo, while those particles with  a semi-major axis  larger  than five times the orbital radius of the main ring are removed from the system. The numerical simulation  has been carried out for 10,000~years. 

Figure~\ref{anelpart}  shows   the  initial  positions of the particles (0~h), and the positions   after 15~hours,   15~days (15~d),   30~days, 1000~days and after  the complete time span of the numerical simulation (10,000~days).  One day corresponds approximately to 3.43  Chariklo's rotation period. The blue circle shows the corotation semi-major axis, the green and the red circles show the location of the    1:2 and 1:3 spin-orbit resonances, respectively.

Under the effects of Chariklo's non-axisymmetric gravity field,  the particles are removed from the inner region  through collisions with Chariklo and ejections from the system. As can be seen, after 10000~days, Chariklo's elongation  clears a  region up to  the location of the   1:2 spin-orbit resonance (green circle), where the stable region begins. 

This result shows the size of the unstable region for the recent refined physical parameters  of  Chariklo [\citetalias{Morgado2021}]. Comparing our results with those presented by \cite{Sicardy2019}, there is a difference in the location of the inner edge of the stable region. This is mainly due to  the assumed  value of $C_{22}$, which is derived from the  semi-axes ($a$, $b$ and $c$). The  value of $C_{22}$ obtained from the semi-axes  given by  \citetalias{Morgado2021} is almost half of the value assumed by \cite{Sicardy2019}, which can explain the decreasing in the unstable region.

\subsection{Poincar\'e Surface of Section}

It is known that the system has an integral of  motion, the  Jacobi constant ($C_{j}$). This conserved quantity, given by \citep {HU2004}
\begin{equation}
\label{eq-Cj}
C_{j} = \omega ^{2}(x^2+y^2)+2U(x,y)-\dot{x}^{2}-\dot{y}^{2}
\end{equation}
 is used in the construction of the Poincar\'e surface of section. In order to unequivocally determine a particle in the phase space it is necessary to have four elements: two for the position $(x,y)$ and two for the velocity $(\dot x, \dot y)$. Fixing a value for the Jacobi constant $C_j$ it becomes necessary to have just three of the four elements, for example, $x$, $y$ and $\dot x$. Defining the section as being  $y=0$, all points of the trajectory that cross this section in a given direction ($\dot{y}_{0}<0$) can be plotted in the plane $(x,\dot{x})$,  producing the Poincar\'e surface of section.
Several  works  also used the Poincar\' e surface of section technique to study a two-body problem system with a central body with non-spherical distribution of mass \citep{SCHEERES1996, Jiang2016, feng2017, Borderes2018, Winter2019,madeira2021}.

This is a numerical procedure where the equations of motion (\ref{eq:movx}) and (\ref{eq:movy}) were integrated using the Bulirsh-Stoer integrator \citep{Bulirsch1966}. The Newton-Raphson method was used to obtain the points of the trajectory that cross the section defined by $y=0$, with an error of at most $10^{-13}$. For each Jacobi constant, $C_{j}$, we considered between 20 and 30 initial conditions. The choice of initial conditions is carried out  into two stages:

\begin{itemize}
    \item firstly, we distribute a set of initial conditions equally spaced on the O$x$ axis, with $x=0$ and $\dot{y}>0$. The $\dot{y}$ velocity is calculated using Equation~\ref{eq-Cj}. These first initial conditions can generate the chaotic regions, the family of the first type orbits and the families of resonant orbits that have islands passing  through the $\dot{x}=0$ axis on the Poincar\'e surface of section.
    \item For the resonant orbits that do not have islands crossing the $\dot{x}=0$ axis in the Poincar\'e surface of section, a second distribution of initial conditions equally spaced in $x \neq 0$ is needed. This distribution occurs in the region where the resonance is found in the Poincar\'e surface of section.
\end{itemize}

The results of a surface of section are geometrically interpreted in a simple way, since periodic orbits produce a number of fixed  points, while quasi-periodic orbits generate islands (closed curves) around the fixed points. Points spread over an  area of the section are identified as chaotic trajectories.

As a reminder,  we consider the  Jacobi ellipsoid shape model proposed by \citetalias{Morgado2021}, see Table \ref{tab:char} . Since $C_{22}$ is different from zero, resonances between the spin of Chariklo and the orbital motion of the particle appear in the system.

\begin{figure*}
\centering
\includegraphics[width=\columnwidth]{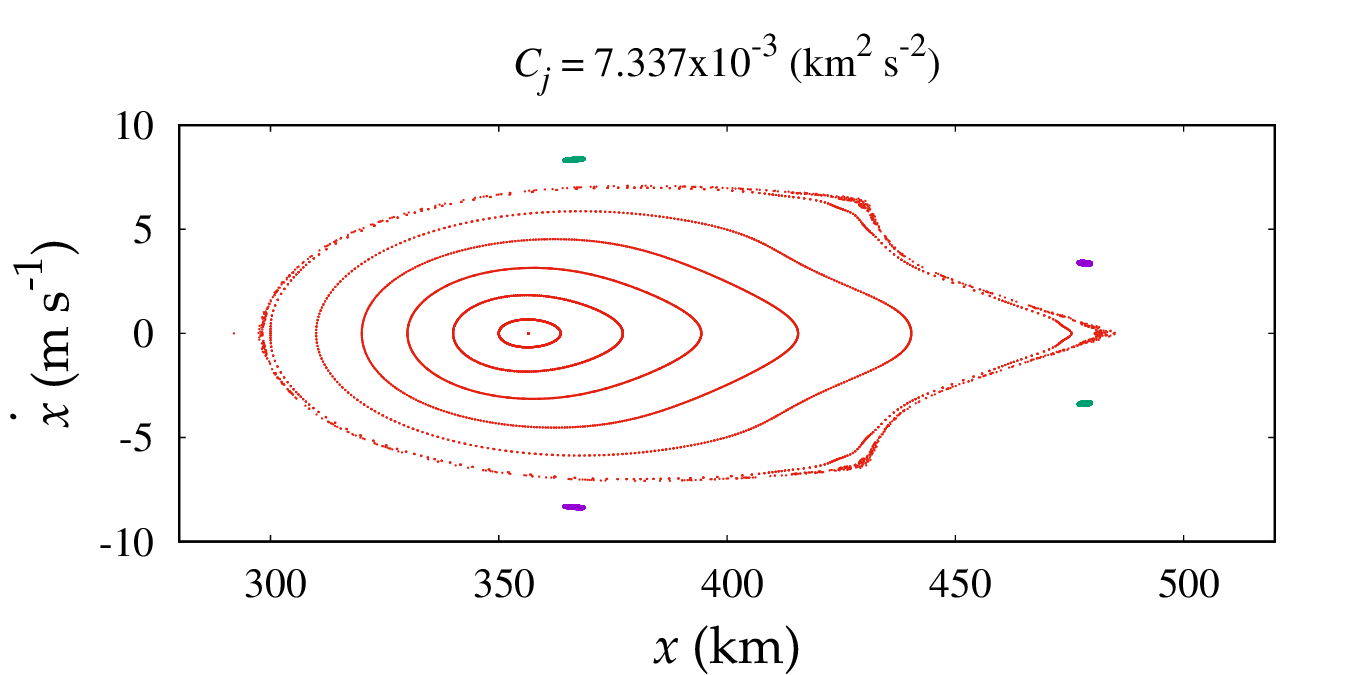}
\includegraphics[width=\columnwidth]{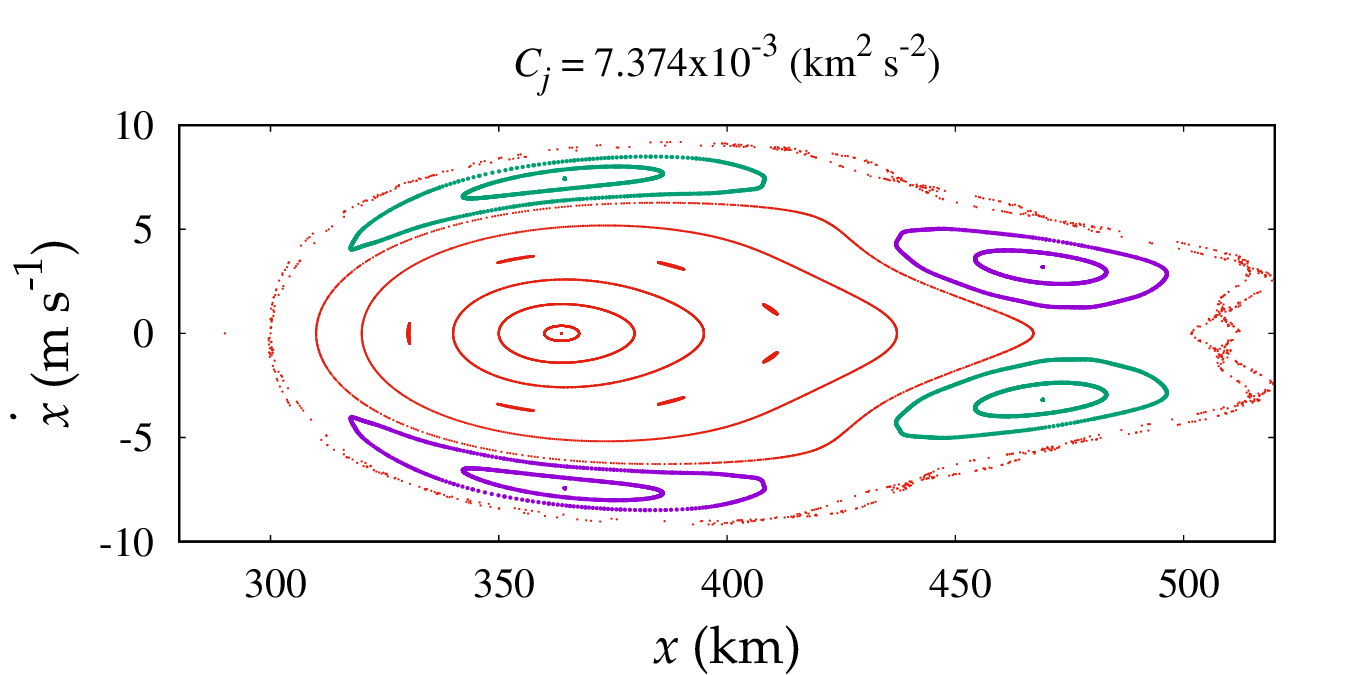}
\includegraphics[width=\columnwidth]{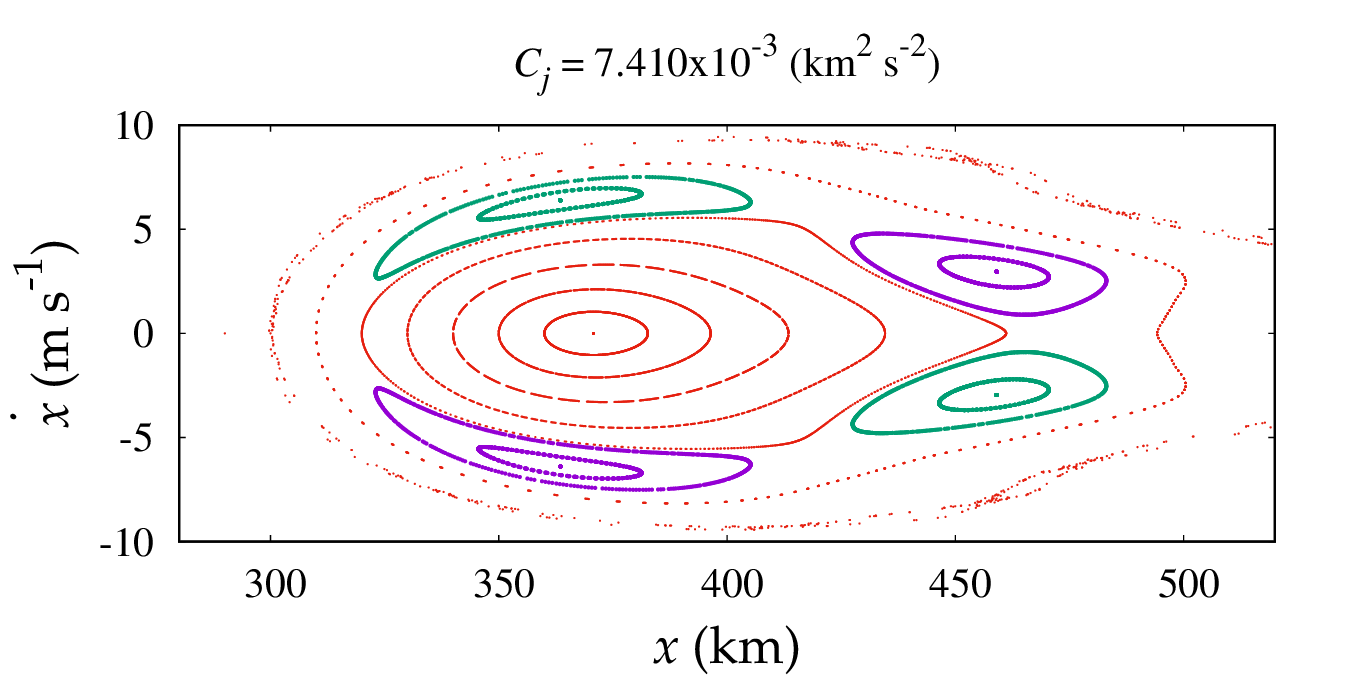}
\includegraphics[width=\columnwidth]{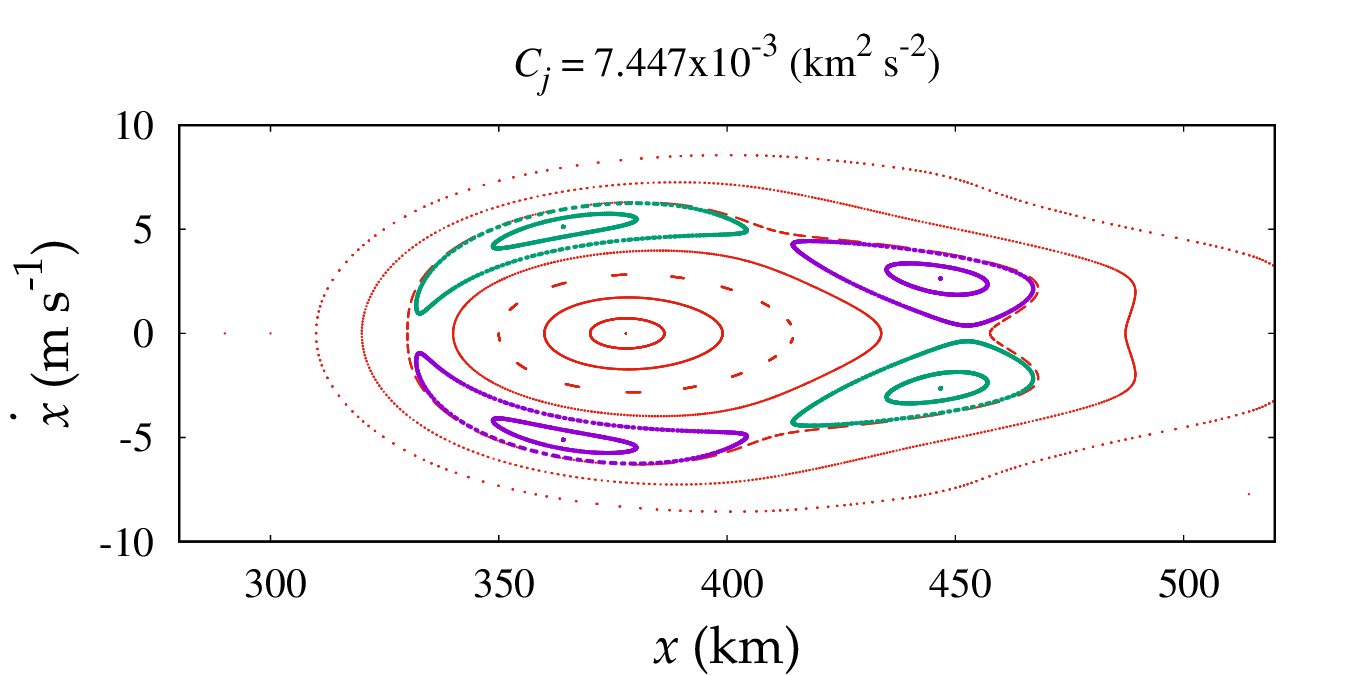}
\includegraphics[width=\columnwidth]{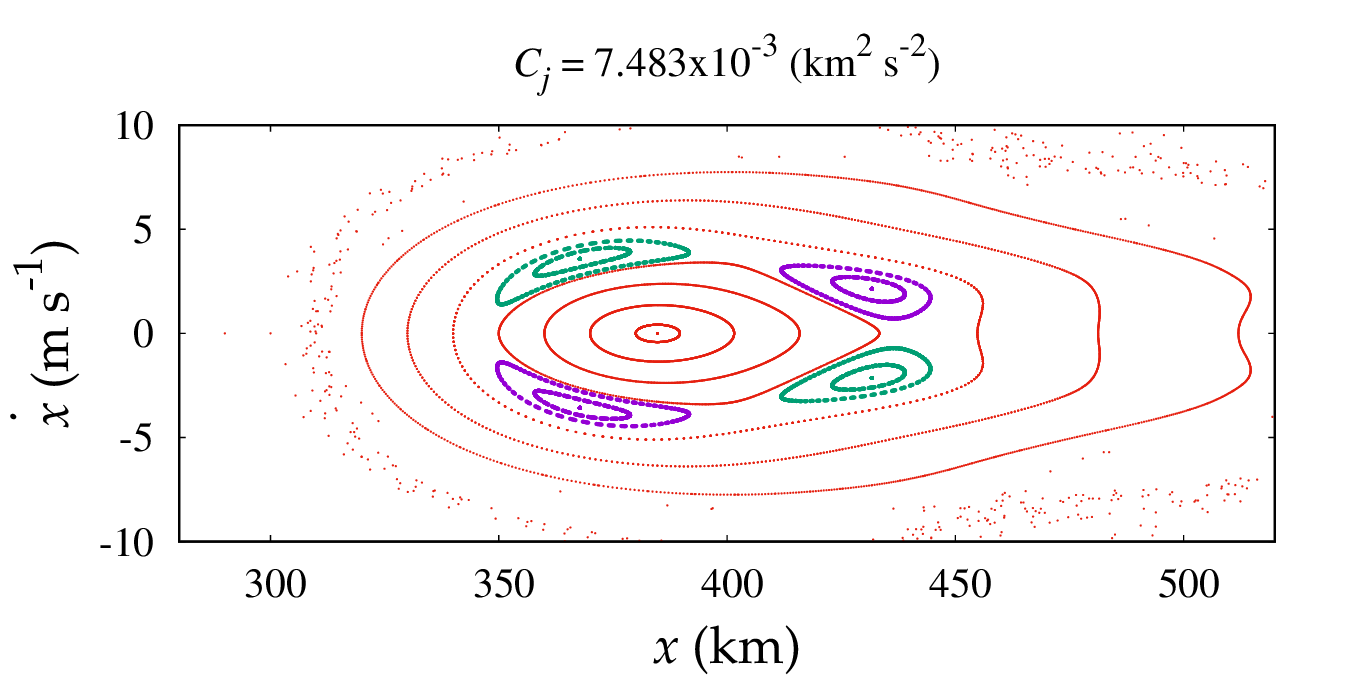}
\includegraphics[width=\columnwidth]{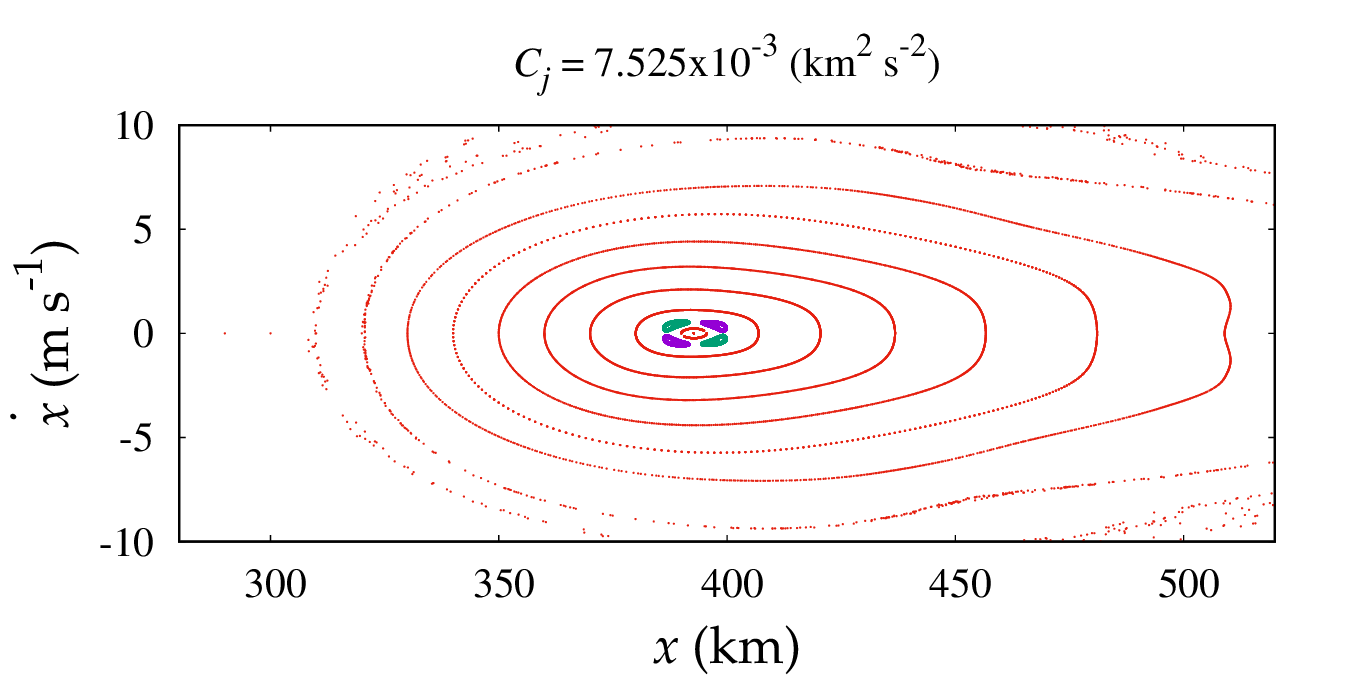}
\caption{Set of Poincar\' e surface of sections covering the whole 1:3 resonance evolution. 
There are two families of 1:3 resonant periodic orbits indicated with different colours, purple and green.
A family of first kind periodic orbits is responsible for the  quasi-periodic orbits shown as  red  islands.}
\label{fig_PSS}
\end{figure*}

The  1:3 spin-orbit resonance is of particular interest once it is close to the location of the ring \citep{Leiva2017}. Through the  Poincar\' e surface of sections, we explored a wide range of $C_{j}$ values in order to identify the islands associated with the 1:3 resonance. These islands exist for $7.337\times 10^{-3}\, {\rm km}^2{\rm s}^{-2}<C_{j}<7.525\times 10^{-3}\, {\rm km}^2{\rm s}^{-2}$. The 1:3 resonance is shown in the plots of Figure~\ref{fig_PSS} (in green and purple colours). 

It is important to notice  that, formally, this 1:3 resonance is a fourth order 2:6 resonance, since the potential (Equation~\ref{eq:U}) is invariant under a rotation of $\pi$ \citep{Sicardy2019, Sicardy2020}, which is a doubled resonance that produces a pair of periodic orbits and their associated quasi-periodic orbits. Consequently, there are two pairs of mirrored sets of islands in the Poincar\' e surface of sections (one pair in green and other in purple), since each one of the periodic orbits generates two fixed points with their islands of quasi-periodic orbits around them. From now on we will refer to 1:3 resonance, instead of 2:6 resonance. It is also valid for the 1:2 (2:4) resonance. 

Once this resonance is doubled in the phase space,  a separatrix appears  between the two families of periodic/quasi-periodic orbits, producing a chaotic region whose size  depends on the Jacobi constant value.  

In the Poincar\' e surface of sections shown in Figure~\ref{fig_PSS}, pairs of islands in green and pairs of islands in purple indicate the quasi periodic orbits associated with the 1:3 resonance. For $C_{j}=7.337\times 10^{-3}\, {\rm km}^2{\rm s}^{-2}$ they are  small and distant from the red islands centre. As the value of $C_{j}$ increases their  size grows and gets closer to  the red islands centre. The green and purple islands are bigger for $C_{j}\sim 7.374\times 10^{-3}\, {\rm km}^2{\rm s}^{-2}$, when they are near to the red islands. As the  $C_{j}$ value keeps increasing, the 1:3 resonance islands start to reduce their sizes and get closer to the  red islands centre, being surrounded by red islands ($C_{j}=7.525\times 10^{-3}\, {\rm km}^2{\rm s}^{-2}$). The evolution continues as they approach to the red islands centre, always getting smaller. A family of periodic orbits of the first kind \citep{Poincare1895} is responsible for the red islands. This will be discussed in details in the following section.

\section{Periodic Orbits} \label{periodicorbit}

Traditionally, periodic orbits in the planar, circular, restricted three-body problem have been classified as periodic orbits of the {\it first kind} and of the {\it second kind} \citep{Poincare1895, Szebehely1967}. The so-called {\it resonant periodic orbits} are the periodic orbits of the second kind, whose particles are in eccentric orbits in a mean motion resonance. The less well known are the periodic orbits of the first kind, which are those originated from particles initially in circular orbits in the unperturbed system (simple two-body problem).

Families of periodic orbits of the first kind have being studied in several systems. For example, in the restricted three-body problem, \citet{Broucke1968} considered the Earth-Moon case, \citet{winter1997a} studied in the Sun-Jupiter system, and  \citet{giuliatti2013} found them in the Pluto-Charon system. Now, considering the restricted two-body problem, where the primary is a rotating non-spherical triaxial body, \citet{Borderes2018} and \citet{Winter2019} showed  examples of both kinds of periodic orbits.

From the set of Poincar\' e surface of sections shown in Figure~\ref{fig_PSS}, the dynamical structure of the  region where are located the rings of Chariklo is determined by two families of  1:3 resonant periodic orbits (in purple and in green colours) and by a family of periodic orbits of the first kind (in red colour). We will explore some features of these periodic orbits.

\begin{figure}
\centering
\includegraphics[width=\columnwidth]{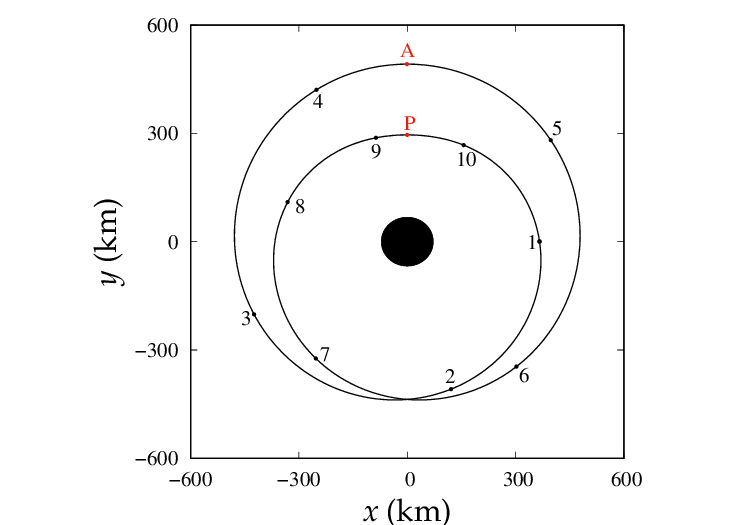}\\
\includegraphics[width=\columnwidth]{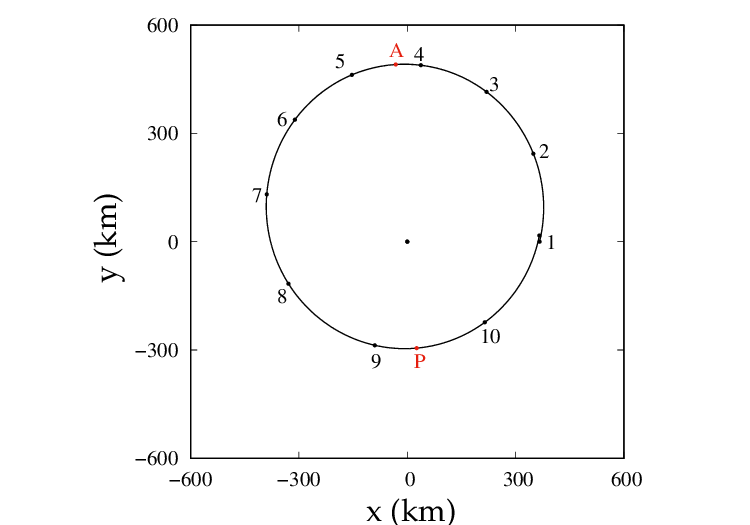}\\
\includegraphics[width=0.78\columnwidth]{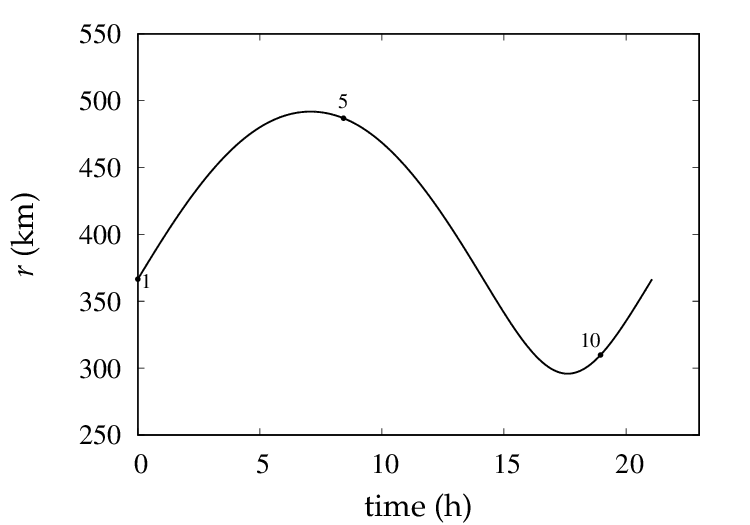}
\caption{Example of a 1:3 resonant periodic orbit whose Jacobi constant value is $C_j=7.337\times 10^{-3}\, {\rm km}^2/{\rm s}^2$ and its period in the rotating frame is $T= 21.01$~h. The points at the centre of the islands, in purple, shown in the Poincar\' e surface of section of Figure \ref{fig_PSS} (first plot) correspond to this periodic orbit. In the top plot is the trajectory in the rotating frame, while the trajectory in the inertial frame is shown in the middle plot. 
The temporal evolution of the orbital radius is shown in the bottom plot. 
The numbers in the top and middle plots show the locations equally spaced in time, indicating the time evolution of the trajectory. The letters $P$ and $A$ indicate the location of the pericentre and the apocentre, respectively.}
\label{fig_orb1}
\end{figure}

\begin{figure}
\centering
\includegraphics[width=\columnwidth]{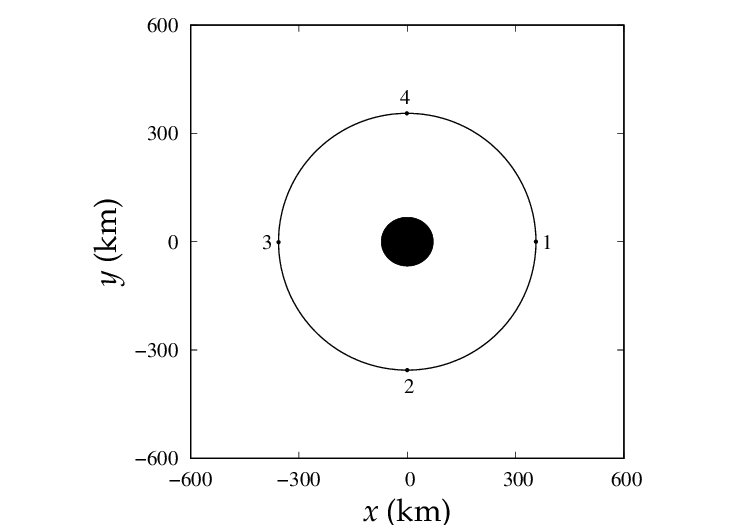}
\includegraphics[width=\columnwidth]{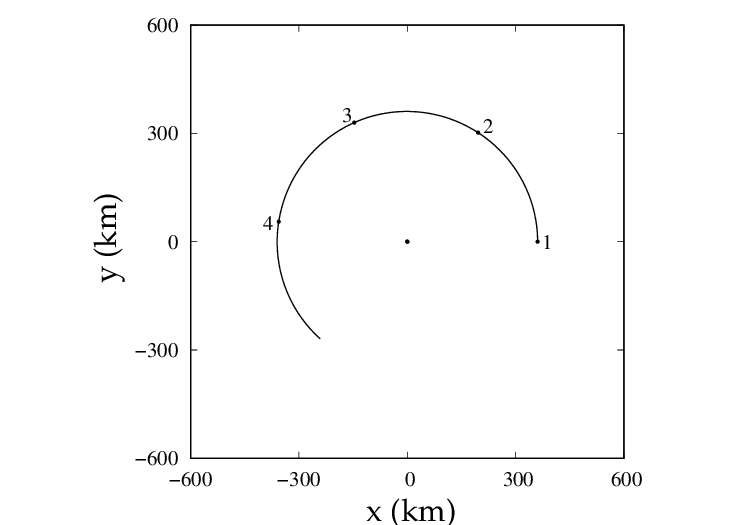}
\includegraphics[width=0.8\columnwidth]{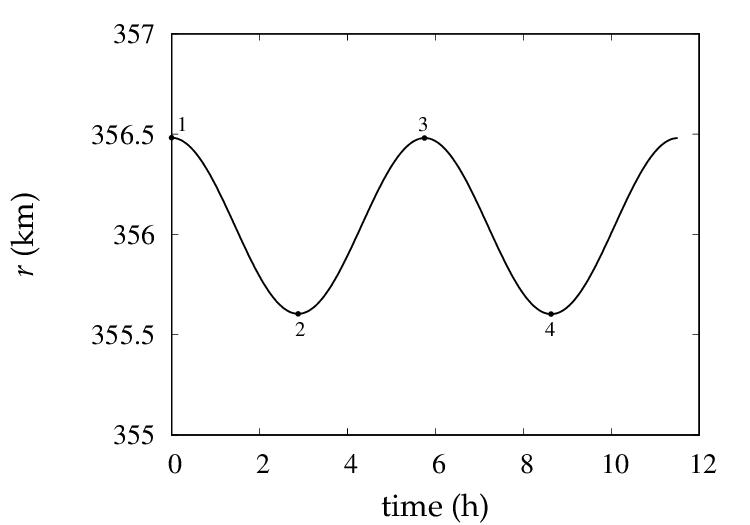}
\caption{Example of a first kind periodic orbit with $C_j=7.337\times 10^{-3}\, {\rm km}^2/{\rm s}^2$ and the period of the periodic orbit in the rotating frame is $T= 11.5$~h. The point at the centre of the islands in red, shown in the Poincar\' e surface of section of Figure \ref{fig_PSS} (first plot), corresponds to this periodic orbit. In the top plot is shown the trajectory in the rotating frame, while the trajectory in the inertial frame is in the middle plot. The temporal evolution of the orbital radius is shown in the bottom plot. The numbers in the top and middle plots show the locations equally spaced in time, indicating the time evolution of the trajectory.}
\label{fig_orb2}
\end{figure}

Since the 1:3 resonance is doubled, we selected just one of the resonant periodic orbits to study. The other orbit is a mirrored image of this one. Considering the Jacobi constant $C_j=7.337\times 10^{-3}\, {\rm km}^2{\rm s}^{-2}$, in Figure \ref{fig_orb1} is shown a 1:3 resonant periodic orbit. This is the periodic orbit shown in the Poincar\' e surface of section of Figure \ref{fig_PSS} (first plot), which corresponds to the points at the centre of the islands shown in purple.   

The top plot shows the trajectory in the rotating frame. The numbers indicate the time evolution of the trajectories and show the locations equally spaced in time. In the rotating frame the trajectory is retrograde and symmetric with respect to the line $x=0$, where are located the pericentre ($P$) and apocentre ($A$) of the trajectory. The period of this periodic orbit is  $T= 21.01$~h, which is very close to three periods of rotation of Chariklo. In the inertial frame, the trajectory is prograde (middle). The temporal evolution of the orbital radii  of the trajectory (bottom) also helps to visualise the trajectory shape. 

The angle of the 1:3 resonance is given by $\phi_{1:3}= \lambda_C - 3\lambda + 2\varpi$, where $\varpi$ and $\lambda$   are the longitude of the pericentre and mean longitude of the particle, respectively, and $\lambda_C$ is the orientation angle of Chariklo. For all resonant trajectories shown in Figure \ref{fig_PSS},  the resonant angle oscillates around $\pi$ (orbits in purple) or $-\pi$ (orbits in green).
  
In Figure \ref{fig_orb2} is presented an example of a first kind periodic orbit, with $C_j=7.337\times 10^{-3}\, {\rm km}^2{\rm s}^{-2}$.
Through a careful analysis, we verified that the trajectory of a particle in the rotating frame always follows a shape similar to the shape of Chariklo. The closest points of the trajectory (indicated by numbers 2 and 4) are aligned with the short axis of Chariklo, while the furthest points of the trajectory  (indicated by numbers 1 and 3) are aligned with its long axis.

The period of this periodic orbit is $T= 11.5$~h, a little more than 1.5 times the spin period of Chariklo. Note that, at the same time it completes one period in the rotating frame, the trajectory completed only a little  more than half of its orbit around Chariklo in the inertial frame. The temporal evolution of the radial distance (bottom plot of Figure \ref{fig_orb2}) clearly shows that the trajectory has a pair of minima and a pair of maxima. This trajectory is not a usual Keplerian ellipse, with the central body at one of the foci. Actually, the trajectory is like an ellipse with the central body at its centre.   

A comparison between the radial amplitudes of oscillation of this periodic orbit of the first kind (bottom plot of Figure \ref{fig_orb2}) and the resonant periodic orbit given in the  bottom plot of Figure \ref{fig_orb1} shows a huge difference. Note that the resonant periodic orbit shows a radial oscillation of more than 190~km while the periodic orbit of the first kind oscillates less than 1~km. Such difference will be analysed for the whole set of periodic orbits in the next section. 

\begin{figure}
\centering
\includegraphics[width=1.0\columnwidth]{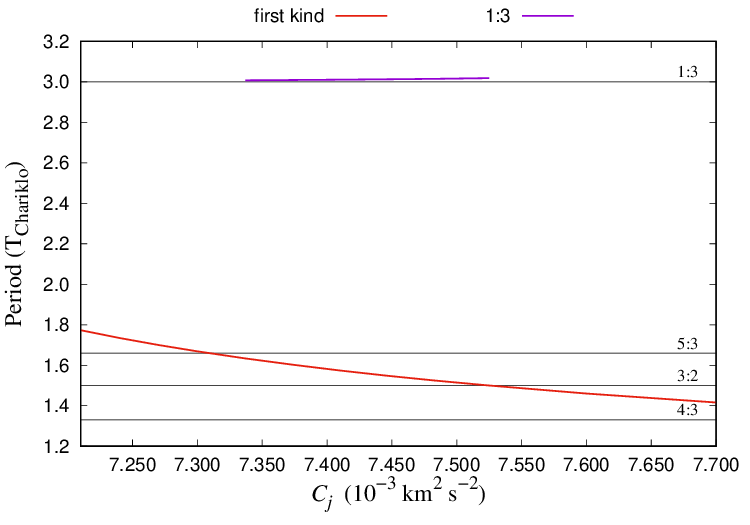}
\caption{Period of the periodic orbits in the rotating frame as a function of their Jacobi constant, $C_j$. 
In red are the periods for the first kind periodic orbits, while in purple are the values for the 1:3 resonant periodic orbits. 
Some periods  that are commensurable with Chariklo's spin period ($T_{\rm Chariklo}$), in the rotating frame, are indicated by the black straight lines.
}
\label{fig_rtd}
\end{figure}

A striking difference between a periodic orbit of the first kind and a resonant periodic orbit is that the period (in the rotating frame) 
of a periodic orbit of the first kind varies significantly, in a range that might cross several values that are commensurable without becoming resonant. The period of the periodic orbits in the rotating frame as a function of their Jacobi constant, $C_j$, is shown in Figure \ref{fig_rtd}.
Note that the resonant periodic orbits 1:3 (in purple) exist only nearby the period commensurable with the spin period of Chariklo, while the period of the periodic orbits of the first kind covers a wide range of values (red),  crossing some periods that are commensurable  with the spin period of Chariklo.

\section{Location of the Rings} \label{locationrings}
Following the approach developed by \citet{Winter2019}, we look for a correlation between the locations and sizes of the rings of Chariklo  and the locations of the stable regions associated with the periodic orbits analysed in the last section.

As seen in Figure \ref{fig_orb2}, in the rotating frame, the first kind periodic orbits have an ellipsoidal shape with Chariklo at its centre, with a radial extent going from a minimum ($r_{\rm min}$) to a maximum ($r_{\rm max}$) radial distance from Chariklo. This same radial extent can be obtained by a Keplerian ellipse with a pair of equivalent semi-major axis and eccentricity ($a_{\rm eq},e_{\rm eq}$), where  $a_{\rm eq}= (r_{\rm min}+r_{\rm max})/2$ and  $e_{\rm eq}= 1-(r_{\rm min}/a_{\rm eq})$ \citep{Winter2019}.

\begin{figure}
\centering
\includegraphics[width=\columnwidth]{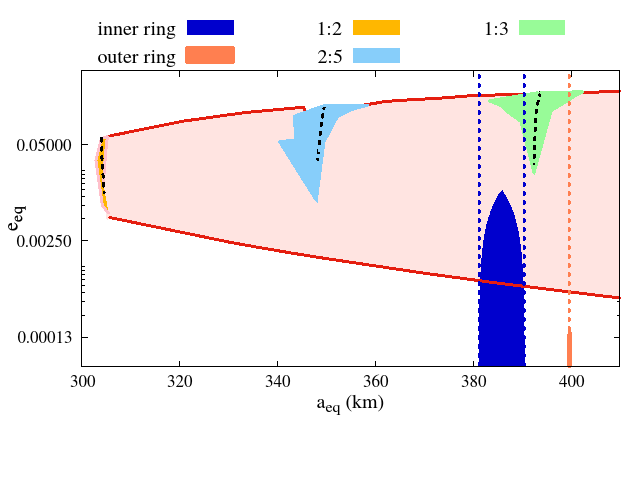}
\includegraphics[width=\columnwidth]{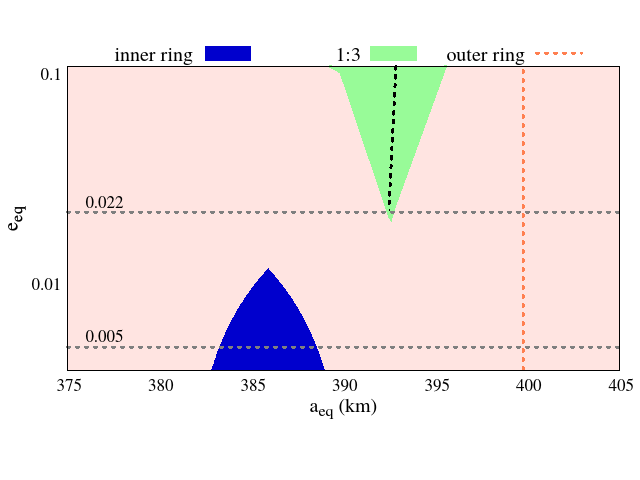}
\caption{(a)  Map of the stable region around Chariklo in $a_{eq}$ versus $e_{eq}$ space. The stable region indicated by the pink colour is due to the first kind orbits. This region is limited by two red lines: bottom, which corresponds to the semi-major axis and eccentricity of the first kind periodic orbits, and upper, which refers to quasi-periodic orbits. The yellow, blue, and green regions are related to 1:2, 2:5, and 1:3 resonances, respectively. Black dashed lines indicate the centres of each resonance, determined through the geometry of the resonant orbits. The dark blue and coral regions show the ranges ($a_{eq}$, $e_{eq}$) that correspond to the locations of the inner (C1R) and outer (C2R)  rings of Chariklo, respectively. The dashed dark blue and coral lines show the edges of the rings.  A zoom  in the  bottom plot (b) shows two horizontal lines which marks the limits of the eccentricities given by \citetalias{Morgado2021}.}
\label{fig_axe}
\end{figure}

Adopting this same idea, each one of the rings of Chariklo can be represented by a set of Keplerian ellipses with equivalent semi-major axis $a_{\rm eq}$ and eccentricity $e_{\rm eq}$, covering the same radial extent of the ring.  Figure~\ref{fig_axe}a shows a comparison of this region covered by these Keplerian ellipses and our results on the stable region around Chariklo, adopting the Jacobi ellipsoid shape model proposed by \citetalias{Morgado2021}. The pink colour refers to  the region of periodic and quasi-periodic orbits of the first kind (red islands in the Poincar\'e surface of sections in Figure \ref{fig_PSS}). This region is bounded by two red lines: bottom, which corresponds to the semi-major axis  and eccentricity of the first kind periodic orbits, and upper, which refers to the quasi-periodic orbits of largest libration, for each value of $C_j$. The yellow, blue, and green regions correspond to the widths of the 1:2, 2:5, and 1:3 resonances, respectively. Resonance width limits are obtained by calculating the pair ($a_{eq}$, $e_{eq}$) of the quasi-periodic orbits with the largest libration for each value of $C_j$. The black dashed lines indicate the centre of each of these resonances.

Additionally, the dark blue and coral regions indicate the equivalent region covered by C1R and C2R, respectively. The width of each ring was derived from the observations \citepalias{Morgado2021} and they are represented by dashed lines \citepalias[we assumed the upper limit of C1R width provided by][$W=9.1$~km]{Morgado2021}. A particle located  on the inner or  outer edge of the C1R  can assume  any  values of the eccentricity and semi-major axis defined  in the dark blue region which will guarantee that the width of C1R  will be about 9~km \citepalias{Morgado2021}.

The width of the dark blue region changes due to the fact that a ring particle with  a non-zero value of eccentricity needs to be located in a particular semi-major axis in order to keep the width of the ring (shown as dashed lines in dark blue and coral colours representing C1R and C2R, respectively). As the eccentricity of the particle increases, the values of the semi-major axis that this particle can be located decrease. For larger values of $e$ ($ > 0.015)$ the excursions of the  ring particle will lead to a  value larger than the width of the ring.  Figure~\ref{fig_axe}b shows a zoom of the C1R region, with the horizontal dotted lines placing the  limits  of C1R eccentricity obtained by  \citetalias{Morgado2021}, eccentricities equal to 0.005 and 0.022.

From a certain value of the eccentricity {\bf ($0.0006$)}, it can be noted that the dark blue region enters the stable region of the orbits of the first kind, indicating that the C1R particles need to have a minimum value of eccentricity to be located in a stable region. The upper limit of eccentricity (0.022), in turn, is outside the dark blue region. This value was obtained by \citetalias{Morgado2021} by adjusting the observed data from Chariklo centre, corresponding to the upper bound for a 3-sigma confidence level. For 1-sigma confidence level, the upper bound is 0.014, a value pretty similar to the maximum eccentricity in the dark blue region (0.012). Therefore, our equivalent region of stability is consistent with the observed location of C1R. 

Note that if  a  particle on the outer/inner edge of a ring  located  in the 1:3 resonance region, for example, for the values $e = 0.0205$ and  $a=392.55$~km (bottom point of the green area),  the radial excursion of the particle would be very large, about 16~km. 

For C2R the situation is different. Since it has a very small width,  around $0.117$~km \citepalias{Morgado2021}, the  eccentricity of the particles  must be very small ($<0.00013$). However, for   C2R  to be in a stable region (pink region), its eccentricity must be larger  than 0.0006. Consequently, its   width has to be  larger than $\sim 120$~m \citepalias{Morgado2021}. This poses a problem with the stability of this ring.

\section{Chariklo rings} \label{secconfi}

The dynamics of Chariklo rings has been discussed since their discovery, with the similarities of these rings with the narrow rings of Saturn \citep[e.g.: Titan, Maxwell, Huygens ringlets,][]{Colwell2009} and Uranus \citep[e.g.: $\alpha$, $\beta$, $\eta$ rings,][]{French1991} being highlighted by several works. In fact, the most recent data on Chariklo system \citepalias{Morgado2021} show that C1R and Uranus $\epsilon$ ring share the characteristics of being narrow, dense, and eccentric.  These analogies are positive, as they enable us to build on our prior knowledge of narrow rings, being a good start to understanding Chariklo inner ring. 

In Section~\ref{subsec_structure}, we use classic planetary ring theory to extract some quantities about Chariklo rings, while in Section~\ref{subsec_confinement} we estimate the timescale of material removal in the ring, in the absence of confinement. In Section~\ref{subsec_models} we propose some confinement models for the rings.

\subsection{Structure of the rings} \label{subsec_structure}

Data from \citetalias{Morgado2021} show that C1R and C2R are both dense narrow rings, with C1R corresponding to an eccentric structure, while C2R is probably a low eccentricity ring. In the C2R case, the maximum eccentricity of the particles can be estimated as the ring half-width over its radial location. From this calculation we  obtained  $e\sim1.5\times10^{-4}$.

Now, C1R must have a positive eccentricity gradient ($\delta e$) and some kind of alignment of the pericentres of both edges is necessary  in order to maintain its observed eccentric configuration. Assuming the simplest streamline mode for C1R and an initial alignment of the apse in the ring,  the minimum $W_{-}$ and maximum $W_+$ radial widths of the ring are given by \citep{Nicholson1978}
\begin{equation}
W_\mp=\delta a(1\pm q),    
\end{equation}
where $\delta a$ is the semi-major axis width of the ring (or mean width) and $q$ is the dimensionless eccentricity gradient. Taking the limit values obtained by \citetalias{Morgado2021} (Table~\ref{tbinit}), we find $\delta a=6.95$~km and $q=0.31$. From $q\approx a~\delta e/\delta a$ \citep{French1986} we obtain a positive eccentric gradient for the ring of $\delta e=0.0056$.

Observational data from the narrow rings of Uranus and Saturn show that these structures show apse alignment, possibly caused by a combination of internal torques in the ring \citep{Goldreich1979a,Borderies1983,Chiang2000,Mosqueira2002,Papaloizou2005}. When the ring is narrower in periapse than in apoapse (which is true for $\delta e>0$), the self-gravity effects would be increased in the periapse. In this position, the outer half of the ring would be radially pulled by the inner half, resulting in a differential precession that (almost) cancels out the effect caused by the central body. Consequently, the ring would have a rigid precession, as a single entity \citep{Goldreich1979a,Borderies1983}

Additionally, a large amount of collisions occurs at each orbital period in the ring, producing impulses that contribute to the differential precession. Inside the ring, the timescale of precession caused by collisions is much longer than that caused by the central body, and can be disregarded. However, if shepherd satellites are confining the ring, they will generate pressure-induced acceleration in the particles, producing a double-peak profile in the ring \citep{Chiang2000}.

\cite{Melita2020} obtain a double-peak profile for the rings of Chariklo, which is roughly consistent with the W-shape of C1R observed by \citetalias{Morgado2021}. Furthermore, \cite{Melita2020} found that Chariklo rings require surface densities of $\sim 10^2$~kg/m$^2$, one order of magnitude lower than the values estimated by \citetalias{BragaRibas2014}, and a minimum eccentricity gradient of $q=0.01$. This value is smaller than the value obtained by us, indicating that C1R is more eccentric than predicted by the theory.

We get an estimate of the rings surface density by invoking the fact that a collisional disc under the effects of the pressure, self-gravity, and rotation is stable when its Toomre parameter is of the order of unity. Toomre parameter is given by \citep{Toomre1964}:
\begin{equation}
Q=\frac{\Omega_kc_s}{\pi G\Sigma},
\end{equation}
where $\Omega_k$ is the angular frequency, $c_s$ is the dispersion velocity in the disc, $G$ is the gravitational constant, and $\Sigma$ is the surface density.

The dispersion velocity is related to the ring scale height $H$ by $c_s=H\Omega_k$ \citep{Adachi1976}, where we take the latter as a radially dependent function $H=r_{\rm orb}\sin{i}$, where $r_{\rm orb}$ is the orbital radius at Chariklo and $i$ is the inclination of the rings, taken as $i=5\times10^{-4}$~deg. This value was chosen to obtain rings of few meters in height, as proposed by \citetalias{BragaRibas2014}. Taking $Q=1$, we find the following relation for surface density 
\begin{equation}
{\Sigma\approx\frac{\Omega_k^2r_{\rm orb}\sin{i}}{\pi G},} \label{eq_surface}
\end{equation}
getting $\Sigma\approx 110$~kg/m$^2$ for Chariklo rings, which is in agreement with the values obtained by \cite{Melita2020}. 

The surface density for each ring is given in Table~\ref{tbtimes} along with the other values that will be estimated in this section. It should be noted that all quantities presented in Table~\ref{tbtimes} are obtained following classic prescriptions that assume a spherical central body. Therefore, they must be interpreted as very rough estimates, as we are not aware of how these prescriptions are affected by the shape of Chariklo.

\begin{table*}
\centering
\caption{Quantities $\Sigma$, $m_{\rm r}$, $\overline{s}$, $\tau_{\rm dp}$, $\tau_{\rm dp}$ and $\tau_{\rm pr}$ estimated for C1R and C2R.}\label{tbtimes}
\begin{tabular}{lcccl} \hline \hline
& Symbol & C1R & C2R & Comments \\ \hline
Surface density (kg/m$^2$)   &  $\Sigma$ &   118      &       109    &   see Equation~\ref{eq_surface}    \\
Ring mass (kg)   &  $m_{\rm r}$    & ${\rm 2\times 10^{12}}$      &    ${\rm 3\times 10^{10}}$   &  see Equation~\ref{eq_mass}   \\
Mean radius of particles (cm) & $\overline{s}$  &     22     &       140     &  see Equation~\ref{eq_size}       \\
Differential precession timescale (yr) &   $\tau_{\rm dp}$            &     0.001              &     0.06      &   see Equation~\ref{eq_dp}   \\  
Inter-particle collisions timescale (yr)   &   $\tau_{\rm ic}$          &     4000              &     7.3     &  see Equation~\ref{eq_ic}     \\  
Poynting-Robertson timescale (yr)   & $\tau_{\rm pr}$         &       10$^9$      &        10$^8$   & see Equation~\ref{eq_pr} \\ \hline  \hline        \end{tabular}
\end{table*}

With the surface density in hand, we can estimate the ring mass $m_r$ by \citep{Goldreich1979c}
\begin{equation}
{m_r=2\pi r_cW\Sigma,} \label{eq_mass}
\end{equation}
where $r_c$ is the ring's central position and $W$ is its width\footnote{To obtain the quantities shown in Table~\ref{tbtimes}, we assumed the width of C1R as the mean width $\delta a$.}. As a result, we obtain that if each ring was originated from the disruption of an ancient body made of ice, it must have a radius of at least $R_{\rm pb}=780$~m and $200$~m to produce the masses of C1R and C2R, respectively.

Another quantity that can be obtained is the mean radius $\overline{s}$ of the particles in the ring \citep{Goldreich1982} 
\begin{equation}
\overline{s}=\frac{3\Sigma}{4\rho\tau}    \label{eq_size}
\end{equation}
where $\rho$ is the density of the particles assumed to be made of ice ($\rho=10^3$~kg/m$^3$) and $\tau$ is the optical depth. The radius $\overline{s}$ corresponds to the radius of the ring particles if they all had a single size. This quantity will be used as a fiducial value in later relations that require particle radius. We find $\overline{s}=22$~cm and $140$~cm for C1R and C2R, respectively.

Such quantities can be used to evaluate the necessity for confinement, which we will be discussed below. This can be done by estimating the spreading timescales of a ring, which must be equal to or greater than the age of the  Solar System. If the spreading timescale is less than the age of the Solar System,  the ring is either a recent feature or  it is confined by some effect that prevents the  spreading. The first possibility is very unlikely since as it would mean that we are at a privileged moment in the Solar System's history,  making the second possibility the most likely.

\subsection{Ring timescales} \label{subsec_confinement}

Several effects spread an unconstrained ring, such as differential precession, inter-particle collisions, and Poynting-Robertson drag. For each of these effects, we obtain a typical timescale that will be used to assess the need for confinement mechanisms. Due to the lack of data, other effects that may contribute to the  ring spreading are not considered, such as plasma drag, Yarkovsky effects and tidal dissipation.

Differential precession results from the effects of the non-sphericity of the central body on the ring particles. Despite the difficulty of working with the classic osculating orbital elements in a system around a prolate body \citep{Ribeiro2021}, we do a simple estimation of differential precession timescale $\tau_{\rm dp}$ by assuming the apsidal precession of the orbits caused by Chariklo oblateness \citep[modified from][]{Murray2000}
\begin{equation}
{\tau_{\rm dp}=\frac{2\pi}{3J_2\Omega_kR_e^2}\frac{\left(r_c^2-\frac{W^2}{4}\right)^2}{r_cW}} \label{eq_dp}
\end{equation}
where $J_2$ is the first seasonal harmonic ($C_{20}$). For  Chariklo case, we estimate $J_2\approx 0.13$, giving a precession timescale for Chariklo rings of $\sim$days.

The loss of energy due to collisions causes the continuous diffusion of particles, and the inter-particle collisions timescale $\tau_{\rm ic}$ can be estimated as the time for a particle walks cross the ring under the gravitational effect of other ring particles \citep{Brahic1977,Borderies1985,Salmon2010}
\begin{equation}
{\tau_{\rm ic}\approx\frac{\Omega_k}{\tau}\left(\frac{\rho W}{c_s}\right)^2.}    \label{eq_ic}
\end{equation}
We find $\tau_{\rm ic}\sim 10^3$~yr and $\sim 10$~yr for C1R and C2R, respectively.

A ring particle constantly absorbs radiation from the Sun and re-emits part of it. Due to the particle's orbital motion, the re-emission is not isotropic, which gives rise to a drag force that causes the particle's orbital decay. The Poynting-Robertson drag timescale $\tau_{\rm pr}$ for a ring is given by \citep{Goldreich1979c,Burns1979}
\begin{equation}
{\tau_{\rm pr}=\frac{W}{r_c}\frac{4\rho sc^2}{9\Phi} }\label{eq_pr}   
\end{equation}
where $s$ is the particle radius, $c$ is the speed of light and $\Phi$ is the solar density flux at Chariklo. For C1R and C2R, we obtain $\tau_{\rm pr}\sim 10^5\times{\rm s(\mu m)}$~yr and $10^4\times{\rm s(\mu m)}$~yr, respectively. 

When we assume $\overline{s}=s$, we find $\tau_{\rm pr}$ of the order of the age of the  Solar System (Table~\ref{tbtimes}), meaning that the rings could survive in the absence of confinement, when only under the Poynting-Robertson effect. However, in the real ring, where there are particles of various sizes,  micrometre and sub-centimetre particles would be removed on much smaller timescales,  only larger particles would remain in the ring after a few millions of years. By analysing mostly the differential precession and inter-particle collisions, the outer ring would spread in a timescale much shorter than the Solar System age, demonstrating that, in fact, Chariklo rings must be confined by any physical process.

The shape and dynamics of a ring are strongly affected by the mass distribution of the central body \citep{Tiscareno2013}, so it would be natural to assume the non-spherical shape of Chariklo as a possible confinement source for the system.  Chariklo mass distribution is responsible for material depletion in  Chariklo vicinity region, however, the latter does not extend to the ring location, as verified in Section~\ref{locationrings} and also in \cite{Sicardy2019}. 

The simplest known confinement mechanism is the confinement of particles around the Lagrangian points of a satellite, in the context of the  restricted planar 3-body problem \citep{Brown1911}. Such a mechanism actually corresponds to a 1:1 mean motion resonance (MMR), and here we are interested in the horseshoe orbits, which are orbits that surround the satellite's $L_3$, $L_4$, and $L_5$ Lagrangian points. The total radial width $W_{\rm hso}$ of the region where particles are confined in horseshoe orbits is \citep{Weissman1974}
\begin{equation}
W_{\rm hso}=a_s\left(\frac{m_s}{M}\right)^{1/3}  \label{hso}  
\end{equation}
where $a_s$ and $m_s$ are the semi-major axis and mass of the satellite, respectively, and $M$ is the central body mass.

Around the horseshoe orbits,  there is a region where the particles exhibit chaotic behaviour and are lost by collisions. The chaos in the system arises from the overlap of first-order MMRs with the satellite \citep{Wisdom1980}, and may be a possible explanation for the apparent gap between C1R and C2R. The width of the gap $W_{\rm gap}$ corresponds to \citep{Duncan1989}
\begin{equation}
W_{\rm gap}=2.6a_s\left(\frac{m_s}{M}\right)^{2/7}   \label{gap} 
\end{equation}

Another mechanism that can confine an eccentric ring, especially its edges, is the eccentric resonance (ER) with a satellite. This type of resonance corresponds to an e-type MMR and is responsible for  exchanging of angular momentum between satellite and particle. In the case of an isolated particle, the ER reduces to the Lindblad resonance, for which angular momentum variations are responsible for affecting the particle's eccentricity \citep{madeira2020,madeira2022numerical}. When we are dealing with a ring as an entity, the variations on the angular momentum act by balancing the viscous effects resulting from particle collisions. Thus, the ER acts by preventing the segment from spreading.

First-order resonances are stronger than higher-order resonances being more likely to confine the rings. The leading term of the torque of a circular satellite $\mathcal{T}_{\rm ER}$ on particles in an isolated $m+1$:$m$ ER is given by \citep{Goldreich1979b,Longaretti2018}
\begin{equation}
{ \mathcal{T}_{\rm ER}=\mp3.76\left(\frac{m_s}{M}\right)^2\left|\frac{a_s}{x_e}\right|^2\Omega_k^2r_c^4\Sigma,  }
\end{equation}
where the upper and lower signs apply to the case where the satellite is internal and external to the ring particles, respectively, and $x_e$ is the distance between the satellite and the ring edge. To confine the ring segment, the torque must be at least equal to the viscous torque $\mathcal{T}_{\rm vis}$ that is given by \citep{Lissauer1981}
\begin{equation}
{\mathcal{T}_{\rm vis}=3\pi\Sigma\left(\frac{\Omega_kr_c\Sigma}{\rho}\right)^2.}
\end{equation}
and  the minimum satellite mass that confines the ring edge is \citep{Longaretti2018}
\begin{equation}
{m_s=1.58\frac{\Sigma}{a_s\rho}\left|\frac{x_e}{a_s}\right|M}   \label{massedge}
\end{equation}

As we approach the ring, we have the overlap of the first-order ERs with the ring edge,  a satellite in such region would give rise to a chaotic region, as already discussed. The torque produced by a circular satellite in this condition is given by \citep{Goldreich1980,Longaretti2018}
\begin{equation}
\mathcal{T}_{\rm ER,c}=\mp3.35\left(\frac{m_s}{M}\right)^2\left|\frac{a_s}{x_e}\right|^4\Omega_k^2r_c^4\Sigma, 
\end{equation}
we can obtain, by assuming the satellite inside the ring, that the mass for the object to keep open a gap of width $W$ in the ring is \citep{Longaretti2018}
\begin{equation}
m_s=1.19\frac{\Sigma}{a_s\rho}\left(\frac{W}{a_s}\right)^{3/2}M   \label{gap2}
\end{equation}

In the following section, we make use of these confinement mechanisms to analyse and propose confinement models for Chariklo rings.

\subsection{Confinement models} \label{subsec_models}
\subsubsection{Confinement by one pair of shepherd satellites}
\begin{table}
\centering
\caption{Locations of hypothetical satellites that may hold an edge of a Chariklo ring, due to ER. From left to right, we show the semi-major axis $a_s$ of the satellites, $m+1$:$m$ ER which they are involved in, their minimum radius $R_s$, and which edge they confine. We emphasise in slanted style the satellites that can simultaneously confine a C1R edge and a C2R edge.\label{ressos}}
\begin{tabular}{llll} \hline \hline
$a_s$ (km) &   $m+1$:$m$ & $R_s$ (m) & shepherding  \\ \hline
314.7  &  4:3      &  540    &  C1R inner edge  \\
348.2  &  8:7      &  420    &  C1R inner edge  \\
\multirow{2}{*}{\textsl{364.1}\tablefootnote{This configuration is shown by the green dot in Figure~\ref{models}a}}  &  \textsl{16:15} & \textsl{330} & \textsl{C1R inner edge} \\
 & \textsl{8:7} & \textsl{390} & \textsl{C2R inner edge} \\
366.0\tablefootnote{This configuration is shown by the green dot in Figure~\ref{models}c} & 18:17 & 320 & C1R inner edge \\
\multirow{2}{*}{\textsl{368.0}}  &  \textsl{21:20} & \textsl{300} & \textsl{C1R inner edge} \\
 & \textsl{9:8} & \textsl{380} & \textsl{C2R inner edge} \\ \hline
\multirow{2}{*}{\textsl{407.7}\tablefootnote{This configuration is shown by the blue dot in Figure~\ref{models}a}}  &  \textsl{12:13} & \textsl{320} & \textsl{C1R outer edge} \\
 & \textsl{25:26} & \textsl{230} & \textsl{C2R outer edge} \\ 
408.6  &  23:24      &  240    &  C2R outer edge  \\
415.8\tablefootnote{This configuration is shown by the blue dot in Figures~\ref{models}b and \ref{models}c} &  14:15      &  290    &  C2R outer edge  \\
429.3  &  8:9      &  350    &  C2R outer edge  \\
\multirow{2}{*}{\textsl{447.6}}  &  \textsl{4:5} & \textsl{450} & \textsl{C1R outer edge} \\
 & \textsl{5:6} & \textsl{410} & \textsl{C2R outer edge} \\ 
\hline \hline
\end{tabular}
\end{table}

\begin{figure}
\centering
\label{model1}\includegraphics[width=\columnwidth]{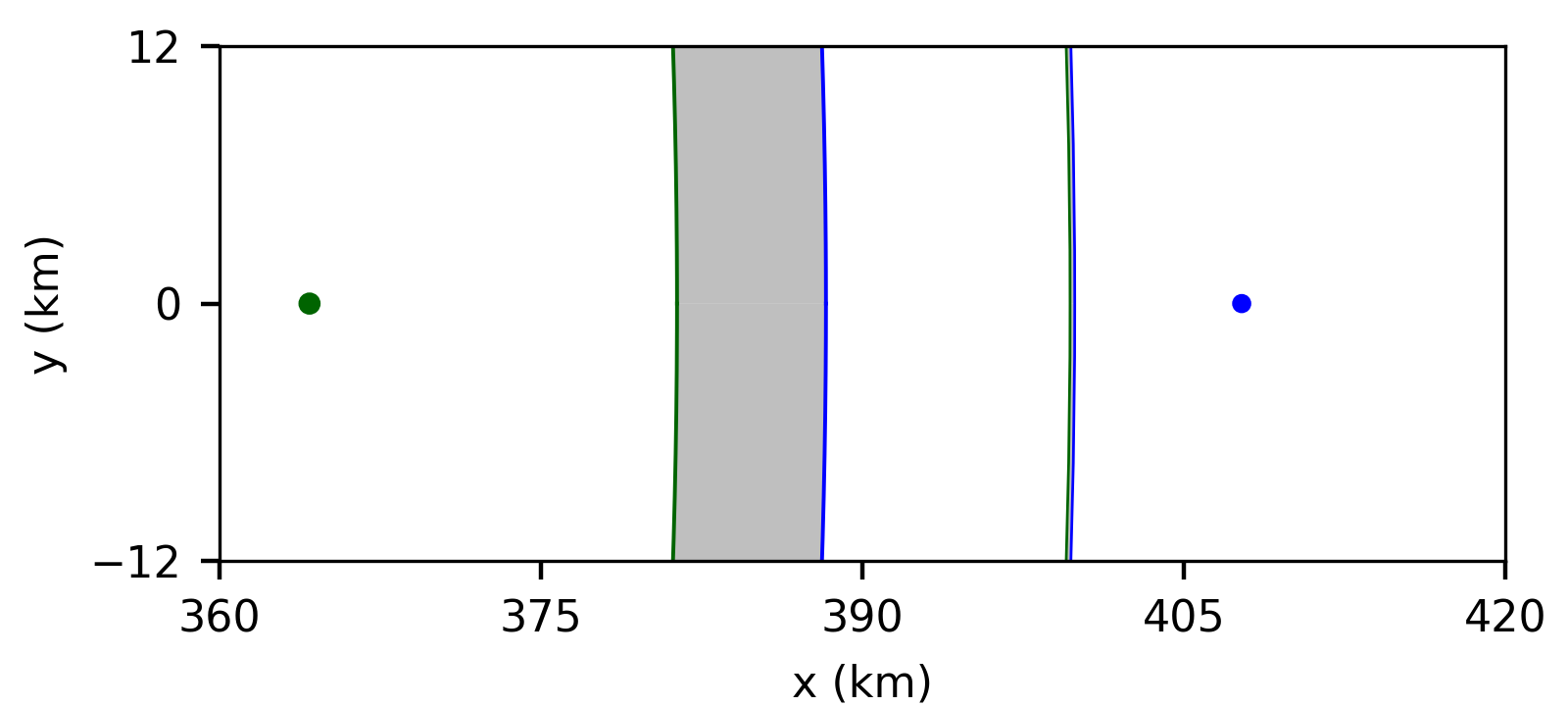}
\label{model2}\includegraphics[width=\columnwidth]{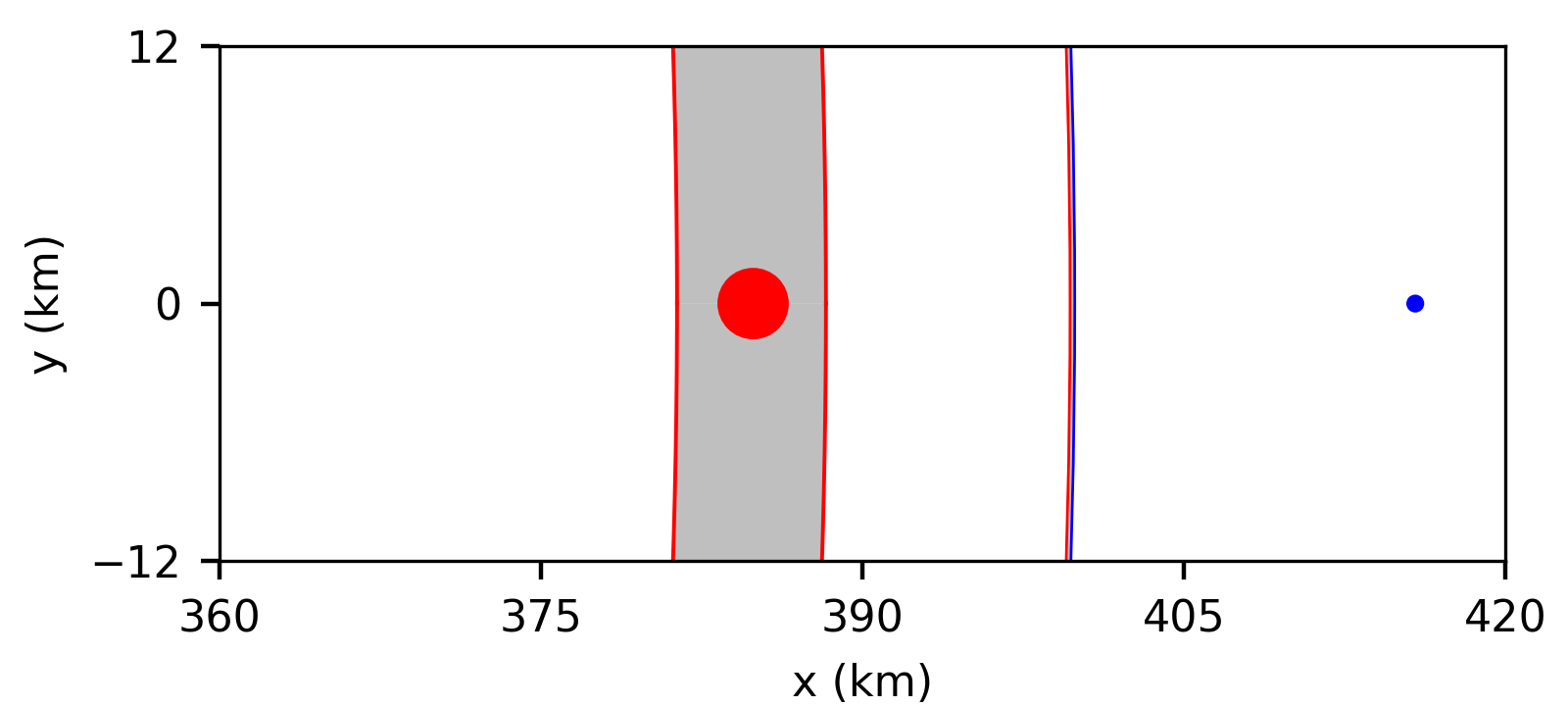}
\label{model3}\includegraphics[width=\columnwidth]{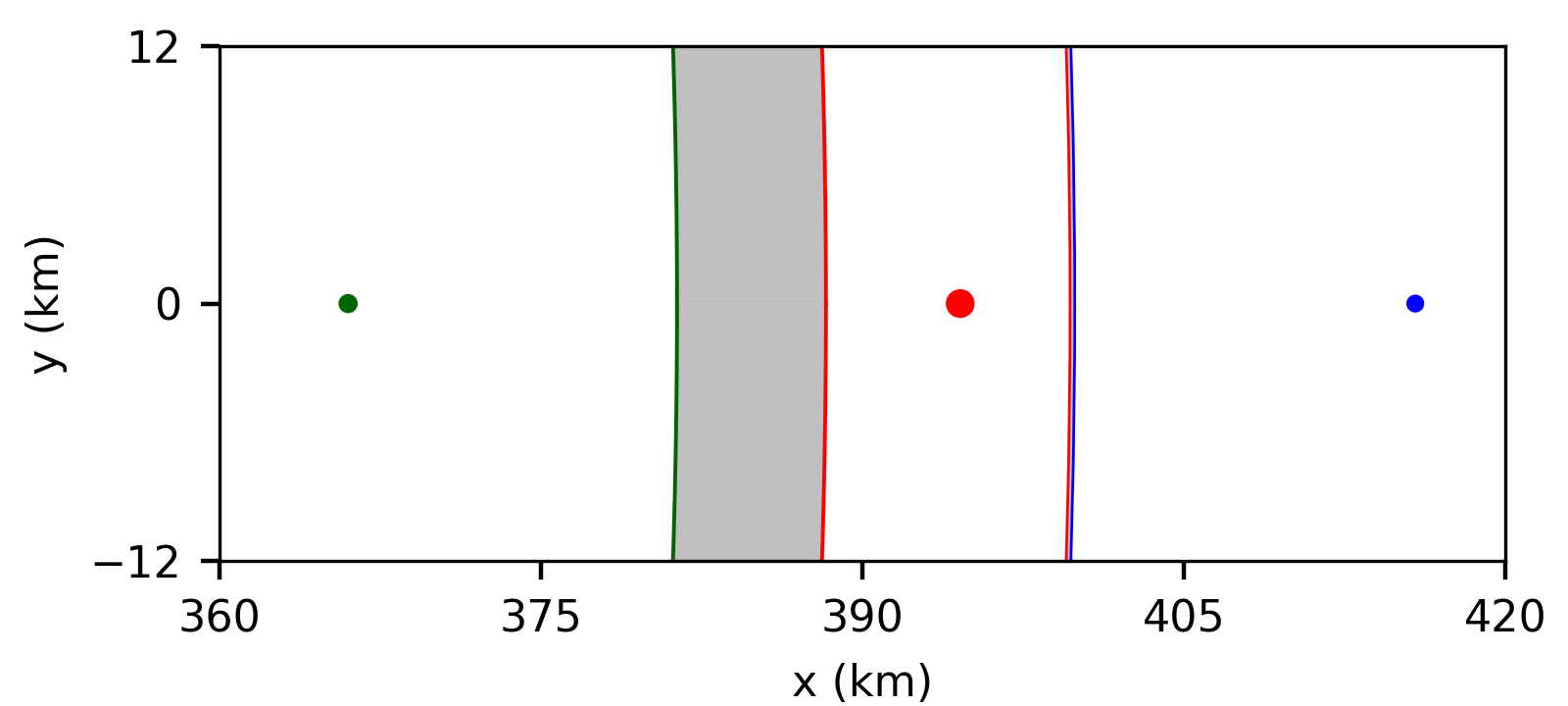}
\caption{Confinement models for Chariklo rings. The top panel corresponds to a model in which the inner/outer edges of both rings are held by the  same inner/outer shepherd satellite. In the middle panel, a satellite confines C1R in its horseshoe region and holds C2R inner edge, while a shepherd satellite confines C2R outer edge. The bottom panel corresponds to a model with the  innermost and outermost edges of the system are confined by a pair of satellites and the gap maintained by a third satellite. Satellites are given by dots with colours matching those of the edges they confine, while the grey regions represent  the width of the  rings. For location and physical radius of  the satellites, refer to Table~\ref{ressos}.}
\label{models}
\end{figure}

Figure~\ref{models}a shows an envisioned confinement model for the rings, in which  the same inner satellite confines the inner edges of both rings, and an outer one confines the outer edges. To obtain the radius and position of the shepherd satellites, we followed an approach similar to \cite{Murray1990} and calculated the location of the $m+1$:$m$ ERs associated with a hypothetical satellite, checking if any matches the edges of the rings. For the inner edges, we look for resonances in the range $1\leq m\leq 30$ and  varied the satellite position from $a_s=160$~km until the C1R inner edge, keeping a semi-major axis interval of 100~m. For the outer edges, we assumed the range of $-30\leq m\leq-1$, with  the satellite position varying by steps of 100~m from the C2R outer edge until $a_s=500$~km. 

The resonance location is obtained numerically following the prescription given in \cite{Sicardy2020} for an ellipsoidal object, with the resonance condition given by 
\begin{equation}
m(\Omega_{k,s}-\Omega_k)=\kappa    
\end{equation}
where $\Omega_{k,s}$ is the angular frequency of the satellite and $\kappa$ is the radial frequency at the edge. After identifying a resonance between satellite and edge, we calculate the satellite's mass using equation~\ref{massedge}. For simplicity, we will assume the satellite with the same bulk density as the ring particles, that is, a satellite made of ice ($\rho=10^3$~kg/m$^3$). For denser materials (e.g. silicates) without porosity, satellites smaller in radius are able to confine the ring, which means that our results correspond to an upper limit on the size of a possible shepherd satellite.

Table~\ref{ressos} shows some possible locations for the shepherd satellites, along with the ER that confines the edge, the minimum satellite radius, and which edge is confined. Due to the proximity between first and second-order resonances for larger values of $m$, there is a possibility that $m+2$:$m$ ERs with the shepherd satellites reside within the ring. These resonances are also responsible for angular momentum exchanges and, if they are inside a ring, they act to excite the particles. Seen that, we present in Table~\ref{ressos} only the cases where the satellite generates a second-order resonance residing within C1R. In slanted style, we emphasise the hypothetical satellites that could confine two edges simultaneously due to discrete ERs, corresponding to cases that could produce the confinement envisioned in this subsection.  In Figure~\ref{models}a, we present only one of the possible satellite configurations capable of confining the rings: the inner edges of C1R and C2R are supported by a $390$~m-sized satellite due to a 16:15 and 8:7 ERs, respectively. The outer edges are confined by a satellite of $320$~m of radius due to 12:13 (C1R) and 25:26 (C2R) ERs. 

Cordelia and Ophelia straddle the  $\epsilon$ ring, but also confine the  $\delta$ (23:22 ER) and $\gamma$ (6:5 ER) rings, respectively \citep{French1991,Nicholson2018}. It is interesting to see that a similar mechanism can occur for Chariklo rings. Based on the numerical simulations of \cite{Hanninen1994,Hanninen1995}, \cite{Goldreich1995} show that extremely narrow rings can be held by a single satellite in eccentric resonance, which we can envision to be the mechanism confining C2R. In this case, we would need only one satellite simultaneously in ER with  one edge of each ring. 

This single-sided shepherding is the mechanism  that confine some ringlets of  the C ring \citep{Lewis2011} and occurs when the satellite torque is sufficient to reverse the angular momentum integrated over the ring streamlines \citep[see][]{Borderies1989}. \cite{sickafoose2019numerical} explored the single-sided shepherding for Chariklo rings and obtained encouraging results indicating that a single satellite might model both rings. 

\cite{Rappaport1998} show that the torque originating from a dense narrow ring can confine the edge of another ring, which can lead us to speculate more complex confinement scenarios, such as satellites confining C2R and the inner edge C1R, while C2R holds the outer edge of C1R. However, such scenarios are only possible under specific conditions, and the confinement by a pair of satellites is a more credible mechanism.

The confinement by only one pair of shepherd satellites has the facility to spare a satellite in comparison to the classic model discussed in Section~\ref{classic}. However, it leaves the system without a mechanism for removing material from the gap. The absence of material between the rings can be explained without additional effects only if the two rings formed independently. Now, if they formed in a single event, an additional mechanism must have acted in the system removing material and helping to define the sharp edges of  both rings.

\subsubsection{Confinement by a co-orbital satellite}

In this section, we rescued the works of \cite{Dermott1979,Dermott1980a} for narrow rings and proposed a model where C1R would be confined to the horseshoe region of a satellite. The interesting point is that such a model has the convenience of explaining the eccentric shape of C1R since the ring will have the same eccentricity as the satellite \citep{Dermott1979}. From equation~\ref{hso}, we obtain that C1R is confined by a satellite with 2~km of radius, shown in Figure~\ref{models}b by the red dot. An external satellite is required to hold C2R outer edge. In the figure, the blue dot corresponds to a satellite holding C2R at 14:15 ER (Table~\ref{ressos}).

A satellite with 2~km of radius would be responsible for generating a chaotic region of  half-width of $15$~km (eq.~\ref{gap}). Given that, the gap has an extension of $13.9$~km, therefore the C2R would be wholly embedded in the chaotic region, which could be an argument to invalidate this model. However, we remind the reader that the recipe to get the chaotic region was developed assuming objects such as mass point. When considering Chariklo as a non-spherical body and the gravitational effects of other nearby satellites, the extension of the chaotic region may change, perhaps allowing a regular motion in the C2R region. In this line of thought, the C2R would reside at the edge of the chaotic region, and therefore a single satellite would confine C1R and produce the gap.

\subsubsection{Classic confinement model} \label{classic}

An already classic confinement model for Chariklo rings was proposed in \citetalias{BragaRibas2014}, in which C1R inner edge and C2R outer edge would be confined by an ER with an inner and outer shepherd satellites, respectively. Additionally, a third satellite would be responsible for the gap between the structures, in analogy, for example, to the satellite Daphnis, responsible for maintaining the Keeler Gap in the A ring of Saturn \citep{Weiss2009}. 

We qualitatively explore this model using the most recent observational data provided in \citetalias{Morgado2021}, first estimating the mass needed for a satellite to hold the gap. Assuming  an object made of ice and located at $a_s=392.9$~km, we obtain from equation~\ref{gap2} a radius of $\sim 150$~m\footnote{For our calculation, we assumed $\Sigma$ as the surface density of C1R.}. 

Equation~\ref{gap2} gives the width of the gap opened by a satellite in a ring  implying that C1R and C2R in the past were a single ring that were separated by this satellite. This is compatible, for example, with a scenario in which the disruption of an object originated a ring of material with the satellite being the largest fragment, or with a scenario whose satellite was formed in situ from material from an ancient ring.

As previously shown in Section~\ref{locationrings}, the 1:3  spin-orbit resonance with Chariklo  is located between the two rings for large values of eccentricity. A satellite trapped in this resonance can open a gap between the rings and prevent them from spreading. If in 1:3 spin-orbit resonance, a satellite in the gap will have a minimum eccentricity of $0.02$ -- and maximum eccentricity of $0.29$ (Figure~\ref{fig_axe}) -- and this, can contribute to C1R eccentricity \citep{colwell2009structure}. Therefore, it is possible that the inner edge of C1R and the outer edge of C2R are being held by shepherd satellites, while a satellite in 1:3 spin-orbit resonance with Chariklo  maintains the gap between the rings.

An example of a system in this confinement model is shown in Figure~\ref{models}c. The innermost satellite (green dot) has $320$~m of radius and confines C1R inner edge (solid green line) due to a 18:17 ER. The outer edge of C2R (solid blue line) in the figure is confined by 14:15 ER with a $290$~m-sized satellite (blue dot), while the C1R outer edge and C2R inner edge (red solid lines) result from the gap caused by a $500$~m-sized satellite (red dot).

The fact that the satellites involved in the confinement have radii  less than one kilometre makes the classic model quite attractive, since a set of reasonable origins can be imagined for these satellites. Some examples are capture or disruption of an ancient satellite. The detection by occultation of angularly tiny objects near a ring is highly unlikely \citep{Sicardy2015}, explaining why these objects were not detected if they actually exist. Finally, the possibility of C1R and C2R having been a single entity in the past is a very appealing point for formation models.

A pragmatic analysis of the models presented here requires more extensive investigations, as it is still unknown how the physical processes discussed here are affected by the ellipsoidal shape of Chariklo. Our intention in this section is not to explain the dynamics of the Chariklo rings but only present some discussions that can be useful for future works on rings around non-axisymmetric bodies.

\section{Discussion} \label{discussion}

The discovery of Chariklo's rings (C1R and C2R) 
brought new insight into planetary ring dynamics. A set of particles orbiting around small objects (compared to the giant planets)  asymmetrically shaped is a new topic to be explored. According to the most recent data \citepalias{Morgado2021} the mean width of C1R is about 7~km (a narrow ring), while its eccentricity is between  0.005 to 0.022. C2R is an even narrower ring, 120~m.  
Our work aims to bring new insights into the dynamics of this unusual ring system. Through a set of numerical simulations and an analysis of periodic and quasi-periodic  orbits, 
we derived the following main results. 
The width of the  unstable region, due to the irregular shape of Chariklo,  is smaller when compared with the results presented in \cite{Sicardy2019}. This difference is caused by the new parameters derived for  the shape of Chariklo.

The presence of gravitational interaction between the ring particles  will cause larger damping due to collisions, making the disk more circular (decreasing in the eccentricity). Although it  probably  will not  change the size of the unstable region. Analysis of  massive bodies with self-gravity is under investigation.

Through a detailed analysis of a sample of Poincar\'e surface of sections we derived the size and location of the stable region in a diagram $a_{eq}$ versus $e_{eq}$. For a given ring width  (9~km for C1R), we computed  values of $e$   and $a$  that a particle located on the inner or outer edge of the ring could assume. This places the  inner ring (C1R) in the stable region if its eccentricity is larger than  $6 \times 10^{-4}$, which corroborates with the recent results. However,  C2R  is located outside the stable region. To be in the stable region,  C2R  needs to have a larger eccentricity value, and consequently a larger width. Therefore, the last  data regarding the width of  C2R needs to be revisited.

Three  confinement mechanisms  are discussed in light of the theory of the known narrow planetary rings of the  giant planets. The classical confinement model seems to be the most suitable. This model requires  three small moonlets to confine the edges of the rings. Two small moonlets, interior to the inner edge of C1R and exterior to the outer edge of C2R, prevent the spreading in  the rings through ERs (mean motion resonance). The gap between the rings would be  opened by a third satellite located in the 1:3 spin-orbit resonance.

Due to the small satellites sizes  required by the confinement models, it is possible to form the shepherd satellites in situ from an old ring or directly by disruption of an older object, which would also give rise to a ring of material. Both scenarios share similarities of requiring an ancient ring from which C1R and C2R were carved due to satellites gravitational effects.

The eccentric and narrow shape of Chariklo rings seems to indicate that the system around the Centaur is more complex than we know, probably hosting shepherd satellites. 
%Any model that intends to explain the formation of the Chariklo system needs to encompass the possible formation of these hypothetical satellites, answering whether the satellites are older, contemporary or younger to the rings. 

  During  the review of this paper two   rings (Q1R and Q2R) were discovered by \cite{Morgado2023}  and \cite{Pereira2023}  around the trans-Neptunian object   (5000) Quaoar located at  43.3~au. Besides the rings, a  satellite, Weywot, orbits Quaoar at  about 24 radii of the central body. Both rings are well located   interior to the orbit of Weywot. Q1R, at 7.4 radii from Quaoar,  is a dense and irregular ring, resembling   the clumpy F~ring of Saturn \citep{Morgado2023}. The intriguing fact regarding this system is that  Quaoar's rings are outside to  its  Roche limit, where dense rings are expected to accrete into satellites. \cite{Morgado2023} claimed that collisions  may keep the  ring even outside the  Roche limit.

More data on rings around Centaurs  are needed to assess whether Chariklo is a rule or an exception among this class of objects, what will allow us to trace the plausibility of the mechanisms discussed.  In fact, new data regarding  different ring systems  around different primary bodies  will help us  to unravel the dynamics involved in each system. 

\begin{acknowledgements}

We are  grateful to the referee  for
the  comments and  suggestions which helped us to improve the manuscript. 
This study was financed in part by the
Aperfei\c coamento de Pessoal de N\'\i ıvel Superior (CAPES, Finance Code~001), Funda\c c\~ao de Amparo \`a Pesquisa do Estado de S\~ao Paulo (FAPESP, Proc.~2016/24561-0, Proc.~2018/23568-6, and Proc.~2021/07181-7, Proc.~2023/09881-1), and Conselho Nacional de Desenvolvimento Cient\'\i fico e Tecnol\'ogico (CNPq, Proc.~305210/2018-1 and Proc.~313043/2020-5). 

Thanks to the Brazilian Federal Agency for Support and Evaluation of Graduate Education (CAPES), in the scope of the Program CAPES-PrInt, process number~88887.310463/2018-00, International Cooperation Project number~3266.

GM thanks to the Centre National d'Études Spatiales (CNES) and European Research Council (ERC). Numerical computations were partly performed on the S-CAPAD/DANTE platform, IPGP, France.
\end{acknowledgements}

\bibliographystyle{aa} % style aa.bst
\bibliography{ref.bib}

\end{document}